\documentclass{aastex}
\usepackage{emulateapj5}
\usepackage{epsfig}

\makeatletter
\newenvironment{inlinefigure}{%
\def\@captype{figure}%
\noindent\begin{minipage}{0.999\linewidth}\begin{center}}
{\end{center}\end{minipage}\smallskip}
\makeatother

\newcommand{\m}{\mbox}
\newcommand{\Ms}{\m{$\,M_\odot$}}

\newcommand{\Msyr}{\m{$\,M_\odot\,$yr$^{-1}$}}

\begin{document}

\shorttitle{Low- and Intermediate-Mass X-Ray Binaries}

\shortauthors{PODSIADLOWSKI, RAPPAPORT, \& PFAHL}

\submitted{Submitted to ApJ}

\title{Evolutionary Binary Sequences for Low- and Intermediate-Mass
X-Ray Binaries}
\author{Ph.\ Podsiadlowski}
\affil{Oxford University, Oxford, OX1 3RH, U.K.; \\
podsi@astro.ox.ac.uk}
\authoremail{podsi@astro.ox.ac.uk}

\and

\author{S. Rappaport and E. Pfahl}
\affil{Department of Physics and Center for Space Research, Massachusetts
Institute of Technology;\\
sar@mit.edu, pfahl@space.mit.edu}
\authoremail{sar@mit.edu, pfahl@space.mit.edu}

\begin{abstract}
We present the results of a systematic study of the evolution of low- and
intermediate-mass X-ray binaries (LMXBs and IMXBs). Using a standard
Henyey-type stellar-evolution code and a standard
model for binary interactions, we have calculated 100
binary evolution sequences containing a neutron star and a normal-type
companion star, where the initial mass of the secondary ranges from 0.6
to 7\Ms\ and the initial orbital period from $\sim 4\,$hr to $\sim
100\,$d. This grid of models samples the entire range of parameters one is
likely to encounter for LMXBs and IMXBs. The sequences show an enormous
variety of evolutionary histories and outcomes, where different
mass-transfer mechanisms dominate in different phases. Very few sequences
resemble the classical evolution of cataclysmic variables, where
the evolution is driven by magnetic braking and gravitational radiation
alone. Many systems experience a phase of mass transfer on a thermal
timescale and may briefly become detached immediately after this phase
(for the more massive secondaries). In agreement with previous results
(\markcite{Tauris1}Tauris \& Savonije 1999), we find that all sequences
with (sub-)giant donors up to $\sim 2\Ms$ are stable against dynamical
mass transfer. Sequences where the secondary has a radiative envelope
are stable against dynamical mass transfer for initial masses up to
$\sim 4\Ms$. For higher initial masses, they experience a delayed dynamical
instability after a stable phase of mass transfer lasting up to
$\sim 10^6\,$yr. Systems where the initial orbital period
is just below the bifurcation period of $\sim 18\,$hr evolve towards
extremely short orbital periods (as short as $\sim 10\,$min).
For a 1\Ms\ secondary, the initial period range that leads to the
formation of ultracompact systems (with minimum periods less than
$\sim 40\,$min) is 13 to 18\,hr. Since systems that start mass transfer
in this period range are naturally produced as a result of tidal
capture, this may explain the large fraction of ultracompact
LMXBs observed in globular clusters. The implications of this study
for our understanding of the population of X-ray binaries and
the formation of millisecond pulsars are also discussed.

\end{abstract}

\keywords {binaries: close -- stars: neutron -- pulsars: general --
X-rays: stars -- white dwarfs}

\section{Introduction}
Low-mass X-ray binaries (LMXBs) were discovered nearly 40 years ago,
and there are now $\sim 100$ known in the Galaxy.  Based on their
short orbital periods of $\la 10\,$d and the absence of luminous
companion stars, it is generally inferred that the donor stars in these systems
are typically low-mass stars (i.e., $\la 1 M_{\odot}$).  However,
to-date Cyg X-2 provides the only case in which a low mass for the
donor star has actually been confirmed dynamically
(\markcite{Casares}Casares, Charles, \& Kuulkers 1998;
\markcite{Orosz}Orosz \& Kuulkers 1998).  Nonetheless, a fairly
compelling picture of LMXBs has emerged over the years, wherein a
low-mass donor star, of varying evolutionary states, transfers mass
through the inner Lagrange point to a neutron star (\markcite{Lewin}Lewin,
van Paradijs, \& van den Heuvel 1995).  Only relatively
recently, however, has attention been focused on the possibility that
many, or perhaps most, of the current LMXBs descended from systems
with intermediate-mass donor stars (hereafter IMXBs).

It has long been conventional wisdom that, if the donor star in an
X-ray binary is significantly higher in mass than the accreting neutron
star, mass transfer would be unstable on a dynamical timescale, and
therefore such systems would not survive. The first systematic study
which indicated that such a view was too simplistic was carried out
by \markcite{Pylyser1}\markcite{Pylyser2}Pylyser \& Savonije
(1988, 1989), who considered compact binaries with initial donor masses
up to $2\,M_{\odot}$ and initial orbital periods of $\la 2\,$d.
\markcite{Tauris1}Tauris \& Savonije (1999) extended this work to
show that, even if the donor star is a \mbox{(sub-)}giant, dynamical
mass transfer
is avoided provided that the initial donor
mass is $\la 2\,M_{\odot}$.

More recent theoretical work in trying to understand the origin of the
``LMXB'' Cyg X-2 and, in particular, the high intrinsic luminosity of
the donor star indicates that the mass of the donor must originally
have been substantially larger ($\sim 3.5\,M_{\odot}$) than the
current value of $\sim 0.6\Ms$ (\markcite{King1}King \& Ritter
1999; \markcite{Podsi1}Podsiadlowski \& Rappaport 2000).
The case of Cyg X-2 is particularly important since it provides direct
observational evidence that, even when the mass-transfer rate exceeds
the Eddington rate by several orders of magnitude, such
intermediate-mass systems can survive this phase of high mass-transfer  by
ejecting most of the transferred mass and subsequently mimick
LMXBs. Independently, \markcite{Davies}Davies
\& Hansen (1998) have suggested that IMXBs may be the progenitors of recycled
pulsars in globular clusters. All of these recent developments have
led to a resurgence in interest in IMXBs (also see \markcite{Kolb}Kolb
et al.\ 2000; \markcite{Tauris2} Tauris, van den Heuvel, \& Savonije 2000).

In order to approach this problem in a more systematic way, we have
carried out binary stellar evolution calculations which cover a broad
grid of starting binary parameters, specifically the mass of the donor
star, $M_2$, and the orbital period at the start of the mass-transfer phase,
$P_{\rm orb}$.  At fixed $M_2$, the value of initial orbital period
effectively determines the evolutionary state of the donor star.  This
library of models comprises 17 different donor-star masses between
$0.6\, M_{\odot}$ and $7\,M_{\odot}$, and up to 8 different
evolutionary states (or, alternatively, values of $P_{\rm orb}$).  The
initial orbital periods span the range from $\sim 4$ hours to 100
days.  The starting parameter values associated with this library of
models are summarized in Figure~1.  In this figure we show the initial
binary parameters in a Hertzsprung-Russell (H-R) diagram for the
companion star.  Evolutionary tracks for stars of the same mass, but
which evolve as single stars are superposed for reference.  Contours
of constant initial orbital period for the case of Roche lobe overflow
onto a neutron star of $1.4\,M_{\odot}$ are also included.

	In our binary evolution models, once mass transfer has commenced, it
is sustained by either (i) systemic angular-momentum losses (e.g., magnetic
braking or gravitational radiation), or (ii) expansion of the donor star
due to nuclear and/or thermal evolution.  The mass transfer may proceed on
any of the timescales implicit in the mechanisms listed above, or may in
fact proceed on a dynamical timescale under certain conditions.  All of
these are explored in detail in this study.

	During the mass transfer phases, these objects will generally
appear as X-ray sources (possibly LMXBs, IMXBs).  These sources could be
steady or transient, depending on the size and temperature of the accretion
disk and on the mass transfer rate through the disk.  At the end of the
mass-transfer phase, many of these systems will become binary radio pulsars,
wherein the neutron star has been spun up to high rotation rates by the
accretion of matter.

One of the main objectives of this study is to provide a library of models
that covers the whole range of parameters for LMXBs and IMXBs using a
self-consistent set of binary calculations and to discuss the various
physical phenomena encountered in the process. In a subsequent
study (\markcite{Pfahl}Pfahl, Podsiadlowski, \& Rappaport 2001), we will
use this library to study the population of LMXBs and IMXBs as
a whole by integrating them into a binary population synthesis code
and by comparing the results with the observed population.

In \S 2 of this paper we describe in detail the stellar evolution code
and the binary model used in this study. In \S 3 we discuss the
various types of binary sequences encountered and compare them to
previous studies. In \S 4 we consider the end products of this
evolution and present a new case study for the formation of
ultracompact X-ray binaries. Finally in \S 5 and 6, we discuss the
implications of these results for the population of X-ray binaries and
the formation of binary millisecond pulsars.

\section{Binary Calculations}

\subsection{The Stellar-Evolution Code}

All calculations were carried out with an up-to-date, standard
Henyey-type stellar evolution code (\markcite{Kippenhahn}Kippenhahn,
Weigert, \& Hofmeister 1967), which uses OPAL opacities
(\markcite{Rogers}Rogers \& Iglesias 1992) complemented with those
from \markcite{Alexander} Alexander \& Ferguson (1994) at low
temperatures\footnote{The opacity tables
were kindly provided to us by P. P. Eggleton.}. We use solar
metallicity ($Z=0.02$), a mixing-length parameter $\alpha=2$ and
assume 0.25 pressure scale heights of convective overshooting from the
core, following the recent calibration of this parameter by
\markcite{Schroder}Schr\"oder, Pols, \& Eggleton (1997) and
\markcite{Pols}Pols et al.\ (1997).
To include the effects of pressure ionization in the equation of
state, which is important for low-mass stars, we adopted the
thermodynamically self-consistent formalism of
\markcite{Eggleton}Eggleton, Faulkner, \& Flannery (1973) and
calibrated the continuum depression term so that our models for single
stars compare well with the detailed models of \markcite{Baraffe}
Baraffe et al.\ (1998). Our models agree with these models typically
within a few per cent in radius (for masses as low as
$0.1\,M_{\odot}$), although their luminosities may differ by as much
as $\sim 20$ per cent.  This may be due, in part, to the fact that we
required a helium abundance of 0.295 for $\alpha=2$ to produce a good
solar model at the present age of the Sun, as compared to their value
of 0.282 for $\alpha=1.9$.  Our single-star models become fully
convective at a mass of $0.351\,M_{\odot}$.

\subsection{The Binary-Evolution Code}

Each of the binaries initially consists of a neutron-star primary with an
initial mass $M_1 = 1.4\,M_{\odot}$ and a normal-type secondary of
mass $M_2$.  The effective radius of the Roche lobe, $R_{\rm L}$, is
calculated with the formula of \markcite{Eggleton2}Eggleton (1983),
\begin{equation}
R_{\rm L} = a\, {0.49\,q^{-2/3}\over 0.6\,q^{-2/3} + \ln (1+q^{-1/3})},
\end{equation}
where $a$ is the orbital separation and $q=M_1/M_2$ the mass ratio
of the binary components. To calculate the mass-transfer rate, $\dot{M}$,
we adopted the prescription of \markcite{Ritter}Ritter (1988),
\begin{equation}
\dot{M} = \dot{M}_0\,e^{R-R_{\rm L}\over H_{\rm p}},
\end{equation}
where $R$ is the radius of the secondary and $H_{\rm p}$ the pressure
scale height at its surface. The constant $\dot{M}_0$ is calculated
according to the model of Ritter (1988). The solution of this equation
requires an iteration in the stellar models. We follow the
method described by Braun (1997), which uses a combined
secant/bisection method (the Brent method; see, e.g., Press et al.\ 1992).

Angular-momentum loss due to gravitational radiation is calculated
according to the standard formula (\markcite{Landau}Landau \& Lifshitz
1959; \markcite{Faulkner}Faulkner 1971),
\begin{equation}
{d\,\ln J_{\rm GR}\over dt} = -{32\over 5}\,{G^3\,c^5}{M_1 M_2 (M_1+M_2)\over
a^4},
\end{equation}
where $G$ and $c$ are the gravitational constant and vacuum speed of
light, respectively.  To calculate the angular-momentum loss due to
magnetic braking, we use the prescription of \markcite{Rappaport1}Rappaport,
Verbunt, \& Joss (1993) (their eq.~36 with $\gamma =4$), which is based on the
magnetic-braking law of \markcite{Verbunt}Verbunt \& Zwaan (1981),
\begin{equation}
{d\,J_{MB}\over dt} = -3.8\times 10^{-30}\,M_2\,R^4\,\omega^3
\mbox{\ \ \rm dyn\,\,cm}.
\end{equation}
In this equation $\omega$ is the angular rotation frequency of the
secondary, assumed to be synchronized with the orbit. We only include
full magnetic braking if the secondary has a sizable convective
envelope, taken to be at least $2\,\%$ in mass (see also Pylyser \& Savonije
1988). For secondaries with convective envelopes smaller than $2\,\%$,
we reduce the efficiency of magnetic braking by an {\it ad hoc} factor
$\exp\{-0.02/q_{\rm conv} + 1\}$, where $q_{\rm conv}$ is the
fractional mass of the convective envelope. We also assume that
magnetic braking stops when the secondaries become fully convective
(\markcite{Rappaport1}Rappaport et al.\ 1983;
\markcite{Spruit}Spruit \& Ritter 1983).

We do not follow the tidal evolution before the onset of mass transfer
(see, e.g., \markcite{Witte} Witte \& Savonije 2001), but start our
binary sequences assuming that the systems have already circularized
when the secondaries are close to filling their Roche lobes; to be
precise we start our calculations when the mass transfer rate as given
by equation~(2) is $\sim 10^{-14}\,M_{\odot}\,$yr$^{-1}$.

For each sequence, we need to specify what fraction, $\beta$,
of the mass lost by the donor is accreted by the neutron star
and the specific angular momentum of any matter that is lost from the
system. We scale the latter with the specific orbital angular momentum
of the neutron star, i.e., assume that the angular momentum loss
due to mass loss from the system is given by
\begin{equation}
{d\,J_{ML}\over dt} = - \alpha\,(1-\beta)\,a_1^2\,\omega\,\dot{M},
\end{equation}
where $\alpha$ is an adjustable parameter and $a_1$ the orbital radius
of the neutron star. The change of the orbital separation due
to the systemic mass loss alone can be calculated analytically
according to
\begin{equation}
{a^1\over a^0} = \left({M_2^1\over M_2^0}\right)^{C_1}\,
\left({M_1^1\over M_1^0}\right)^{C_2}\,\left({M_1^1+M_2^1\over
M_1^0 + M_2^0}\right)^{C_3},
\end{equation}
where superscripts 0 indicate initial values and superscripts
1 final values and where the exponents are given by
\begin{eqnarray}
C_1&=&-2\nonumber\\
C_2&=&-2-2\alpha\,(1-\beta)/\beta\\
C_3&=&1-2\alpha\nonumber.
\end{eqnarray}
When $\beta=0$, equation~(6) has to be replaced by
\begin{eqnarray}
{a^1\over a^0} &=& \left({M_2^1\over M_2^0}\right)^{C_1}\,
\,\left({M_1+M_2^1\over M_1 + M_2^0}\right)^{C_3}\,\nonumber\\
\noalign{\vspace{5pt}}
&&\hspace{1cm}\times
\exp\left\{2\alpha\,\left({M_2^1-M_2^0\over M_1}\right)\right\}.
\end{eqnarray}
In all of our sequences, we take $\alpha$ to be 1, which implicitly
assumes that all the mass lost from the system is lost from the
neighborhood of the neutron star (or its accretion disk), and set
$\beta$, somewhat arbitrarily, equal to 0.5. In addition, we limit the
maximum accretion rate onto the neutron star to the Eddington
accretion rate, taken to be $\dot{M} = 2\times 10^{-8}\Msyr$ and kept
constant throughout each run. In our calculations with relatively
massive secondaries, the mass-transfer rate can exceed the Eddington
accretion rate by many orders of magnitude. Most of this excess mass
must be lost from the system, as the case of Cyg X-2 has
demonstrated. This mass loss may, for example, occur in the form of a
relativistic jet from the accreting neutron star or a
radiation-pressure driven wind from the outer parts of the accretion
disk (see, e.g., \markcite{King2}King \& Begelman 1999).  Evidence for
both of these processes is seen in the X-ray binary SS 433
(\markcite{blundell}Blundell et al.\ 2001), the only system presently
known to be in an extreme super-Eddington mass-transfer phase.

Since the pressure scale height at the surface of the donor is
generally a small fraction of the stellar radius (often as low as
$\sim 10^{-4}\,R$), the calculation of the mass-transfer rate
according to equation~(2) requires that the radius of the star be
calculated to very high precision. To avoid discontinuous changes in
radius and hence $\dot{M}$, it is important that the chemical profile
of the initial star is well resolved and that abrupt changes in the
surface abundances (for example, as a result of dredge-up) are
avoided. To calculate mass loss efficiently, we introduced a moving
mesh in the outermost 5\,\% of the mass of the star. We also assumed
that the outermost $10^{-4}$ of the envelope mass of the donor star
was in thermal equilibrium. This is necessary since in each time step
we typically take off a much larger fraction of the mass of the star
and since the structure variables often change by a large factor in this
outermost layer. It is also justified since the thermal timescale of
this layer is much shorter than any mass-loss timescale encountered in
this study.  (We have extensively tested that our results are not
sensitive to these assumptions, at least for the mass-loss rates
obtained, where we generally limited the maximum mass-loss rate to
$10^{-4}\,M_{\odot}\,$yr$^{-1}$.)

Despite of these precautions, our calculated mass-loss rates are occasionally
subject to numerical oscillations. These tend to be almost negligible
for stars with radiative envelopes (typically less than a few percent),
but can be several 10's of per cent for stars with convective envelopes
and occasionally much larger for evolved giants (in particular during
dredge-up phases). We note that, in all of the plots of $\dot{M}$ presented
in this paper, these oscillations, which do not affect the secular
evolution of the systems, have been averaged out.

\begin{figure*}[t]
\centerline{\epsfig{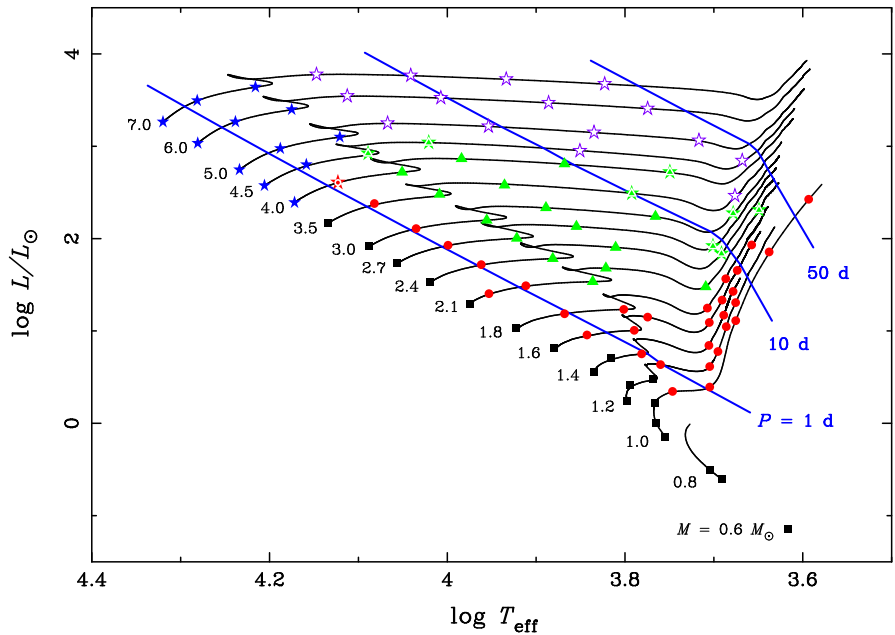}}
\caption{Overview of the initial parameters of
the secondaries in the Hertzsprung-Russell (H-R) diagram for the 100 binary
evolution sequences in the study. The thin solid curves represent
evolutionary tracks of single stars with the masses as indicated.  The
thick solid curves show the orbital periods of systems when the
secondaries start to fill their Roche lobes at the particular
positions along the evolutionary tracks (assuming a mass for the
primary of 1.4\Ms, as appropriate for a neutron star). The symbols
are placed at the initial positions of the secondaries in our sequences,
where the type of symbol indicates the final fate of the systems (squares:
ultracompact systems; circles: He white dwarfs; triangles: HeCO white
dwarfs; filled stars: delayed dynamical mass transfer; unfilled stars:
dynamical mass transfer; in cases where it is not clear whether the
system experiences a dynamical instability, symbols are plotted on top
of a star).}
\label{fig:fig1}
\end{figure*}

\subsection{Tests and Comparisons}

To test our binary evolution code, we chose to calculate the standard
evolution of a cataclysmic variable (CV), initially consisting of a white
dwarf of $0.6\,M_{\odot}$ and a secondary of $0.8\,M_{\odot}$
(here we assumed that all of the mass transferred from the secondary
was lost from the system). In this calculation, the system experienced
a period gap between 2.4 and 3.1\,hr, somewhat smaller than the observed
gap (\markcite{Ritter2}Ritter \& Kolb 1998), but consistent with
previous results for our adopted magnetic-braking law
(\markcite{Rappaport2}Rappaport, Verbunt, \& Joss 1983).
The minimum period in
this calculation was 75\,min, somewhat longer than the minimum
period found in the most detailed studies of CVs (see, e.g.,
\markcite{Kolb2}Kolb \& Baraffe 1999; \markcite{Howell}Howell,
Nelson, \& Rappaport 2001).

We also compared our calculations to other recent similar binary
calculations by a number of different authors, in particular the
calculations by \markcite{Pylyser1}\markcite{Pylyser2}Pylyser \&
Savonije (1988, 1989); \markcite{Han}Han, Tout, \& Eggleton (2000);
\markcite{Langer}Langer et al.\ (2000);
\markcite{Kolb}Kolb et al.\ (2000);
\markcite{Tauris2} Tauris et al.\ (2000). For comparable models, we
generally find excellent agreement between our calculations and the
calculations of these authors. The only significant discrepancy to
note is the early case B calculation for Cyg X-2 by Kolb et al.\
(2000), where the secondary has an initial mass of 3.5\Ms\ and has just
evolved off the main sequence at the beginning of mass transfer. While
our early-case B model (see Podsiadlowski \& Rappaport 2000) is in
excellent agreement with a similar calculation by Tauris et al.\
(2000), in the Kolb et al.\ model, the early super-Eddington phase is
much longer, and as a consequence the mass-transfer rate in the
subsequent slower phase about an order of magnitude lower than in our
model. We do not know the reason for this discrepancy, whether it has
to do with the treatment of mass loss at these very high rates ($\sim
10^{-5}\,M_{\odot}\,$yr$^{-1}$) or whether it is caused by differences
in the structure of the initial models (U. Kolb 2000, private
communication). For example Kolb et al.\ (2000) do not include
convective overshooting in their calculations; this produces a
different chemical profile just outside the hydrogen-exhausted core,
which may affect the evolution of the secondary (the evolutionary
track of their secondary in the H-R diagram is indeed quite
different).  Until this discrepancy is resolved, we note that there is
some uncertainty in the modeling of these systems with extreme
mass-transfer rates.

\begin{figure*}[t]
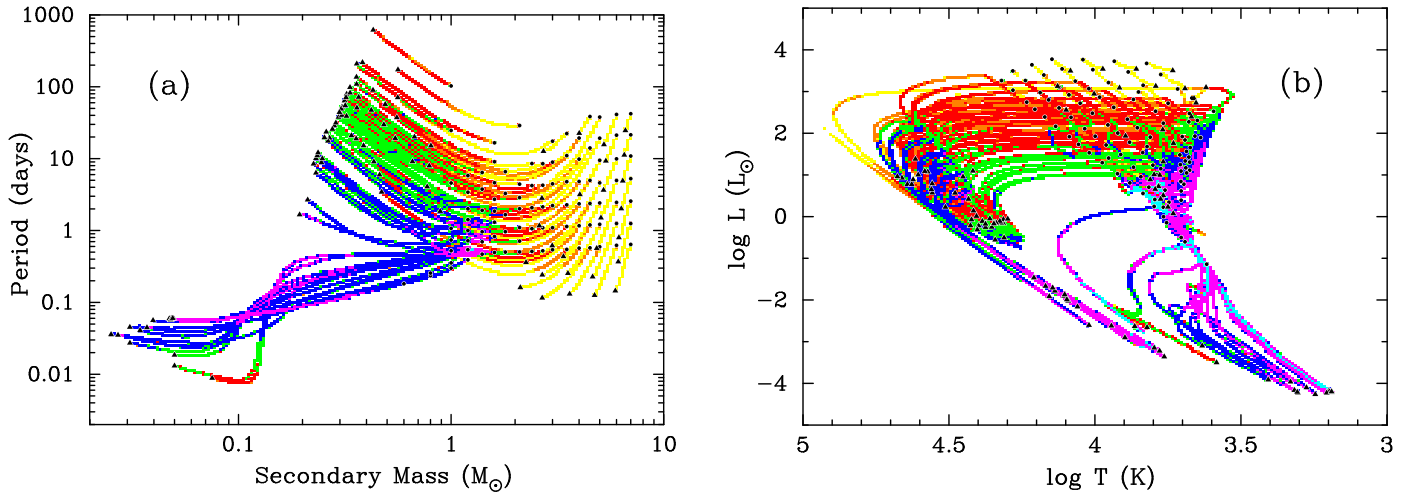

\begin{minipage}[t]{0.48\linewidth}
\centerline{\epsfig{file=fig2a.ps,height=\linewidth,angle=-90}}
\label{fig:2a}
\end{minipage}
\hfill
\begin{minipage}[t]{0.48\linewidth}
\centerline{\epsfig{file=fig2b.ps,height=\linewidth,angle=-90}}
\end{minipage}
\caption{Time-weighted evolutionary tracks for the 100 binary
sequences in the secondary mass -- orbital period plane (panel a) and
for the secondaries in the H-R diagram (panel b). The color of the
tracks indicate how much time systems spend in a particular
rectangular pixel in the diagrams (from short to long: yellow, orange,
red, green, blue, magenta, cyan). The minimum time displayed was
chosen to be 100\,yr, and the maximum time in each of the panels is
$\sim 9.5\,$Gyr. The 7 colors are distributed evenly in $\log t$
between these times. Circles and triangles mark the starting and final
points in the sequences, respectively.}
\label{fig:fig2b}
\end{figure*}

\section{Results of Binary Calculations}

Altogether we carried out 100 binary stellar evolution calculations
with initial secondary masses ranging from $0.6\,M_{\odot}$ to
$7\,M_{\odot}$ and covering, in a fairly uniform manner, all
evolutionary stages likely to be encountered for LMXBs/IMXBs, with
orbital periods from 4\,hr to 100\,d (see Fig.~1). In Figure~2 we
present the evolutionary tracks of these calculations both in a
secondary mass -- orbital period diagram ($\log M_2\,$--$\,\log P_{\rm
orb}$; Fig.~2a) and in a traditional Hertzsprung-Russell (H-R) diagram
(Fig.~2b), where
the color coding indicates how much time a system spends in a
particular region of the diagrams. What these figures do not show very well,
however, is the actual variety in these sequences. Some 70 of the 100
sequences are qualitatively different with respect to the importance
and the order of different mass-transfer driving mechanisms, the
occurrence of detached phases, the final end products, etc. Indeed
there are very few sequences that resemble the classical CV evolution
where mass-transfer is driven solely by gravitational radiation and
magnetic braking.  Instead of presenting all of these sequences in
detail, we will discuss the various physical phenomena encountered
and illustrate them with particular evolutionary
sequences. In the appendix we present the key characteristics of each
sequence in tabular form.

As Figure~2 shows, the sequences can be broadly divided
into three classes: (1) and (2) systems evolving to long periods and short
periods, respectively, and (3) more massive systems experiencing dynamical mass
transfer and spiral-in (the short yellow tracks). The systems evolving
towards short and long orbital periods are separated by the well-known
bifurcation period that has been studied by several authors
in the past (see, in particular, \markcite{Tutukov}Tutukov et al.\ 1985;
\markcite{Pylyser1}Pylyser \& Savonije 1988; \markcite{Ermga1} Ergma 1996;
\markcite{Ergma2}Ergma \& Sarna 1996). For our binary model,
the bifurcation period occurs around 18\,hr for a 1\Ms\ model (see
\S 4.2 for a systematic case study). This implies that all 1\Ms\ models that
start mass transfer on or just off the main sequence evolve towards short
periods, while for the more massive secondaries only relatively
unevolved secondaries do so (see Fig.~1), in agreement with
the findings of \markcite{Pylyser1}Pylyser \& Savonije (1988). However,
the value of the bifurcation period and the behavior of the evolutionary tracks
near the bifurcation period is very sensitive to the model assumptions,
in particular the magnetic-braking law (\markcite{Pylyser1}Pylyser \&
Savonije 1988 and \S~4.2) and the assumptions about mass loss from the system
(\markcite{Ergma1}Ergma 1996; \markcite{Ergma2}Ergma \& Sarna 1996).
Because of the strong divergence of tracks below and about the
bifurcation period, one would expect very few systems with final
orbital periods near the bifurcation period (Pylyser \& Savonije 1988)
unless a system started its evolution very close to it initially
(Ergma 1996).

Figure~2 also shows that, for the more massive systems, the initial evolution
is very rapid. As a direct consequence, very few systems should be observable
in this early rapid phase, and X-ray binaries are most
likely to have a relatively low-mass secondary when they are
observed at the present epoch, even if they had a much more massive
companion initially.

\subsection{Low-Mass Models and the Role of Magnetic Braking}

If the secondary is a relatively unevolved low-mass star initially
(with mass $\la 1\Ms$), the only important mechanisms driving
mass transfer are systemic angular-momentum losses due to
magnetic braking and gravitational radiation. This type of evolution
is similar to the classical evolution of CVs. The systems
evolve towards shorter periods, may experience
a period gap when magnetic braking stops being effective
(when the secondary becomes fully convective) and ultimately reach
a minimum period just before hydrogen burning is extinguished
(\markcite{Paczynski}Paczy\'nski \& Sienkiewicz 1981;
\markcite{Rappaport3}Rappaport, Joss, \& Webbink 1982). Beyond the period
minimum (which depends on the evolutionary stage of the initial model),
the secondaries follow the mass-radius relation for degenerate
stars and the systems will expand, driven by gravitational
radiation alone.

This classical CV-like evolution is illustrated in Figures~3 and 4 for three
binary sequences with initial secondaries of 1\Ms\ and different
evolutionary stages (at the beginning, the middle and the end of the
main sequence).  Figure~3 shows the evolution of orbital period and
mass-transfer rate as a function of time since the beginning of mass
transfer (the evolutionary tracks of the secondaries in the H-R
diagram are shown in Fig.~4). These calculations serve to illustrate
several points, already found in previous studies (see, in particular,
\markcite{Pylyser2} Pylyser \& Savonije 1989). The maximum mass-transfer
rate is of order a few $10^{-9}\Msyr$, where the more evolved secondaries
experience the lower rates. Indeed, this behavior is also often
found for more massive secondaries, where the somewhat evolved stars
generally tend to be more stable than the unevolved ones. The period gap
for the calculation with the initially unevolved secondary
is substantially smaller (2.8 to 3.1\,hr) than the period gap for a
similar CV calculation where the secondary is a white dwarf of 0.6\Ms\
(2.4 to 3.1\,hr). While the donor stars become fully convective at
more-or-less
the same orbital period, they have different masses, 0.336\Ms\
and 0.273\Ms, respectively, since the donor in the LMXB case
is not as much out of thermal equilibrium as in the CV case
(the magnetic-braking timescale is a factor of $\sim 2.2$ longer
in the LMXB case, while the gravitational-radiation timescale
is a factor of $\sim 2.6$ shorter; cf eqs.~3 and 4).

The location and the extent of the period gap decreases for the more evolved
systems and completely disappears for the most evolved one. The reason
is that the more evolved secondaries become fully convective at a
lower mass, which implies a shorter orbital period for the system; but
at shorter orbital periods, the timescales for angular-momentum loss
due to gravitational radiation and magnetic braking become more
comparable, hence producing a smaller gap. Since this type of
evolution is similar to the classical CV evolution, it has the obvious
implication, as emphasized by Pylyser \& Savonije (1989),
that the vast majority of secondaries in CVs have to be essentially
unevolved initially to prevent the appearance of too many systems in
the observed period gap (\markcite{Ritter2}Ritter \& Kolb 1998).

The minimum periods decrease for the more evolved 
systems, again consistent with previous studies (see \S 4.2 for further discussion).
The system that started mass transfer 
\begin{inlinefigure}\label{fig:fig3}
\centerline{\epsfig{file=fig3_c.ps,width=0.9\linewidth}}
\caption{Evolution of orbital period (upper panel)
and mass-transfer rate (lower panel) as a function of time since the
beginning of mass transfer (with arbitrary offset) for three binary
sequences with an initial secondary of 1\Ms\ and three evolutionary
phases (beginning [blue], middle [green] and end of main-sequence
phase [red]; from down to up in the upper panel and up to down in the
lower panel at early times). The least evolved system experiences a
period gap between 2.8 and 3.1\,hr and attains a minimum period of
83\,min, the model in the middle of the main sequence has a period gap
between 2.2 and 2.4\,hr and a period minimum of 80\,min, while the
most evolved system experiences no period gap and has a minimum period
of 48\,min.}
\end{inlinefigure}

\noindent when the secondary had
just completed hydrogen burning in the center attains a minimum
period of 48\,min (note the spike in $\dot{M}$ near the minimum period
in this case).

\begin{figure*}[t]\label{fig:fig4}
\centerline{\epsfig{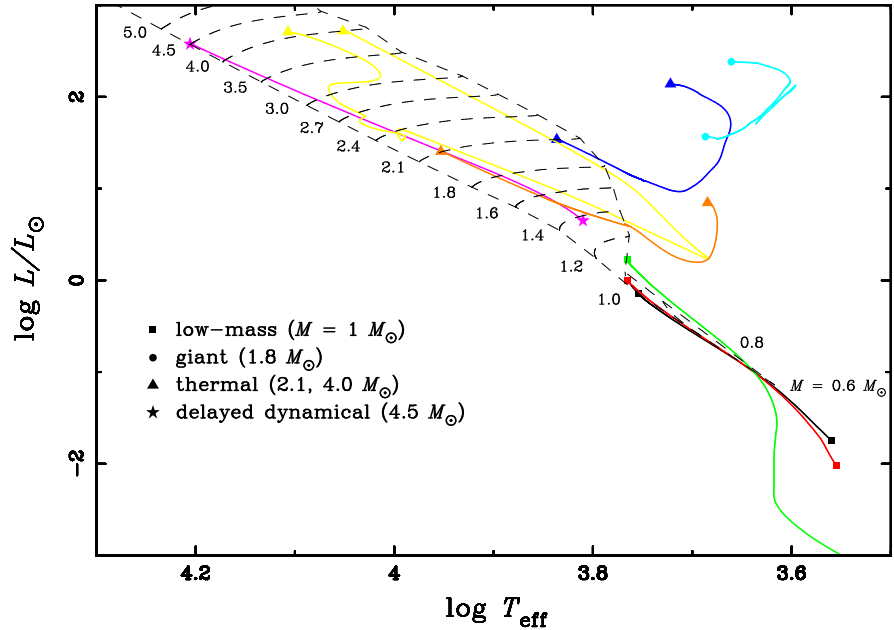}}
\caption{Evolutionary tracks of the secondaries
in the H-R diagram for the illustrative sequences discussed in the
text. The symbols at the end points of the sequences identify the
models (squares: low-mass models [\S 3.1]; triangles: thermal
mass-transfer models [\S 3.2]; circles: evolution on the giant branch
[\S 3.3]; delayed dynamical instability [\S 3.3]).  The sequences are
terminated either near the period minimum (`low-mass'), at the onset of
dynamical instability (`delayed dynamical'), or at the end of mass transfer
(`giant' and `thermal').  The grid of dashed curves indicates the
position of the main-sequence band and the evolutionary tracks for single
stars of the masses as indicated.}
\end{figure*}

\subsection{Thermal timescale mass transfer}

If the donor is initially more massive than the accretor, the Roche lobe
radius generally shrinks. If this radius is less than the thermal
equilibrium radius of a star of the same mass, the secondary can no longer
stay in thermal equilibrium and mass transfer will proceed on a thermal
timescale or, in more extreme cases, on a  dynamical timescale (see
\S~3.3; for general reviews of thermal timescale mass transfer
see, e.g., \markcite{Paczynski2} Paczy\'nski 1970; \markcite{Ritter3}
Ritter 1996, and for other recent discussions \markcite{Distefano}
Di\,\,Stefano et al.\ 1997; \markcite{Langer}
Langer et al.\ 2000; King et al.\ 2001). It is customary to analyze
the stability of mass transfer in terms of mass--radius exponents
where
\begin{equation}
\xi_{\rm eq} = \left({d\ln R\over d\ln M}\right)_{\rm eq},\,\,\,
\xi_{\rm RL} = \left({d\ln R\over d\ln M}\right)_{\rm RL}\nonumber
\end{equation}
\begin{equation}
\mbox{and\,\,\,}\xi_{\rm ad} = \left({d\ln R\over d\ln M}\right)_{\rm ad}
\end{equation}
define, respectively, the mass--radius exponents for stars in thermal
equilibrium, for the Roche-lobe, and for stars losing mass adiabatically.
If the Roche-lobe radius shrinks more rapidly than the
adiabatic radius (i.e., if $\xi_{\rm RL}>\xi_{\rm ad}$),
then there is no hydrostatic
solution for which the secondary can fill its Roche lobe (as
defined by eq.~2) and mass transfer will proceed on a dynamical
timescale (see \S~3.3). The case where mass transfer is dynamically
stable, but occurs on a thermal timescale is given by the
inequalities  $\xi_{\rm ad} > \xi_{\rm RL} > \xi_{\rm eq}$.

For stars with radiative envelopes,
$\xi_{\rm ad}$ is generally very large initially, and in most situations
much larger than $\xi_{\rm RL}$ ($\xi_{\rm RL}$ generally depends on the
mass ratio and any changes in orbital separation due to the transfer
of mass and systemic mass and angular momentum losses; see, e.g.,
\markcite{Rappaport2}Rappaport et al.\ 1983). To illustrate this,
we show approximate adiabatic mass--radius exponents, $\xi_{\rm ad}$,
in Figure~5 for stars with initial masses from 1.2\,--\,2.2\Ms\
and the corresponding mass--radius relations (these were obtained
by taking mass off these stars at a high constant rate of $10^{-5}\Msyr$).
The large initial values for $\xi_{\rm ad}$ imply that the star
has to lose very little mass to shrink significantly. This simply reflects
the fact that, in radiative stars, a large fraction of the envelope
(in radius) contains very little mass (for example, in an unevolved
2.1\Ms\ star, the outer 40\,\% of the radius contains just 1\,\% of
the mass). Once this low-density, high-entropy layer is lost,
$\xi_{\rm ad}$ drops dramatically to a value of order 1 and ultimately
becomes negative when the convective, flat-entropy core is exposed.
From this stage on, the radius of the star increases with further
mass loss. Since the whole
star expands dramatically in this phase, it will be very underluminous
for its mass (since most of the internal luminosity drives the expansion)
and nuclear burning will ultimately be turned off because of a dramatic
decrease
in the central temperature. As long as $\xi_{\rm ad}$ remains larger than
$\xi_{\rm RL}$, mass transfer remains dynamically stable. The star, which is
undersized for its mass, will expand and try to relax to its equilibrium
radius. It is this relaxation of the star on a thermal timescale that gives
this mode of mass transfer its name.

A characteristic mass-transfer rate for this phase is often defined
by an expression of the form (see, e.g., \markcite{Rappaport4}Rappaport,
Di\,\,Stefano, \& Smith 1994; \markcite{Langer}Langer et al.\ 2000)
\begin{equation}
\dot{M}_{\rm th} = {(M_2^i - M_1^i)\over t_{\rm KH}},
\end{equation}
where $M_2^i$ and $M_1^i$ are the initial masses of the secondary
(the mass donor) and the primary, respectively, and $t_{\rm KH}$
is the Kelvin-Helmholtz timescale of the secondary (i.e., the
thermal timescale of the whole star),
\begin{equation}
t_{\rm KH} \simeq {G M_2^2\over 2 R L},
\end{equation}
\begin{inlinefigure}\label{fig:fig5}
\centerline{\epsfig{file=fig5.ps,width=0.75\linewidth}}
\caption{Approximate adiabatic mass--radius relations
(top panel) and adiabatic mass--radius exponents, i.e., $d\ln R/d\ln M$,
(bottom panel) for unevolved stars of masses 1.2, 1.4, 1.6, 1.8, 2.0,
and 2.3\Ms. These were calculated by assuming a constant mass-loss rate
of $10^{-5}\Msyr$. At this high rate, the entropy profile, $S(M)$, remains
almost unchanged (``adiabatic'').}
\end{inlinefigure}

\noindent where $R$ and $L$ are the radius and the nuclear luminosity of the
secondary. As shown by Langer et al.\ (2000), equation~(11) tends
to overestimate the actual mass-transfer rate by up to an order
of magnitude for solar-metallicity stars. In Figure~6 we present
the binary sequence for a 2.1\Ms\ star that starts to fill its
Roche lobe near the end of the main sequence (when its central
hydrogen abundance was $X_c = 0.096$). Indeed, the maximum mass-transfer
rate of $\sim 2\times 10^{-7}\Msyr$ is about an order of magnitude
lower than what equation~(11) would predict. Since the mass-loss
timescale is much longer than the thermal timescale,
the secondary is only moderately out of thermal equilibrium throughout
the high $\dot{M}$ phase.

\begin{figure*}[t]
\begin{minipage}[b]{0.48\linewidth}
\centerline{\epsfig{file=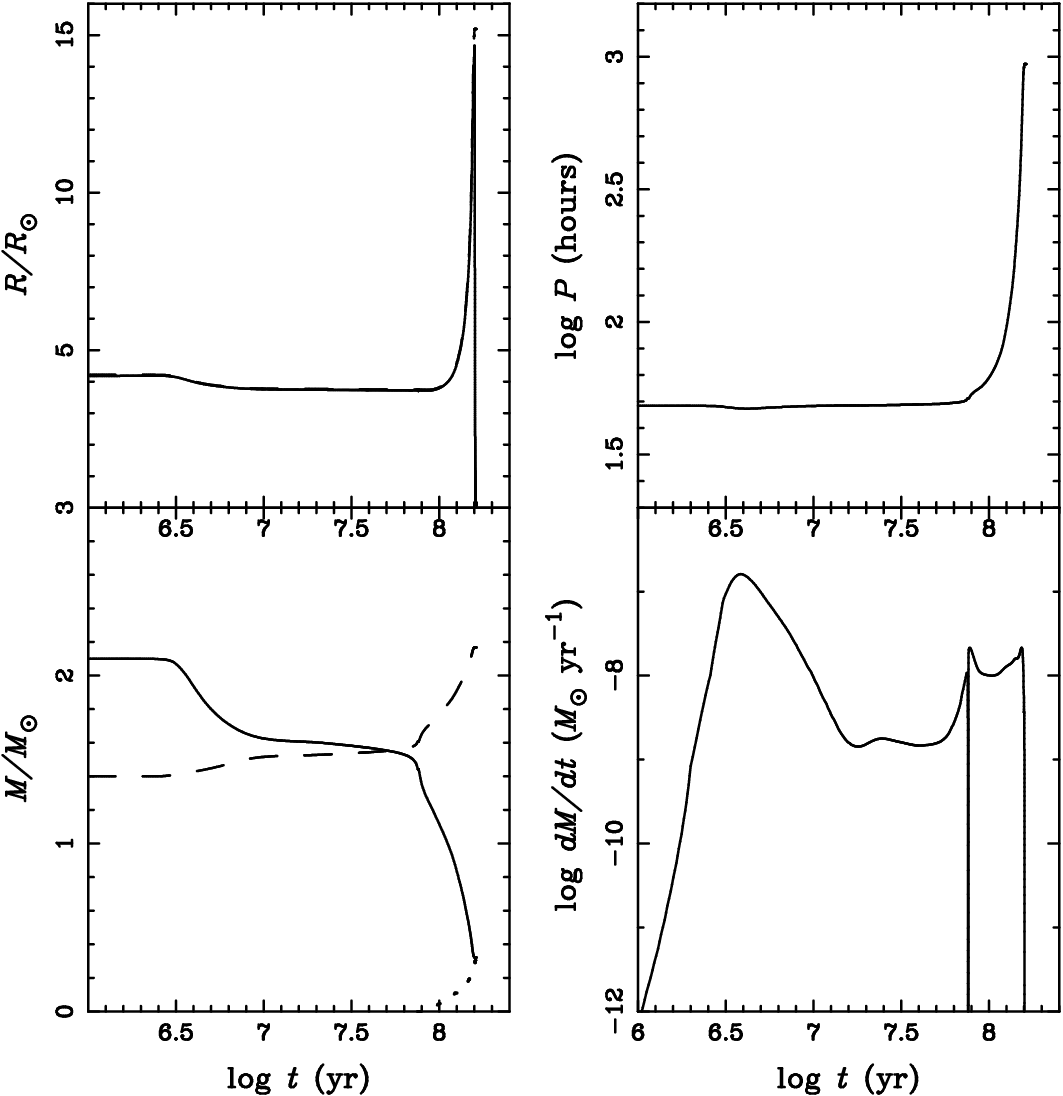,height=\linewidth}}
\caption{Key binary parameters for a binary sequence where the secondary
has an initial mass of 2.1\Ms\ and fills its Roche lobe near
the end of the main sequence (when its core hydrogen abundance
$X_c = 0.1$). Top left:
radius (solid curve) and Roche-lobe radius (dashed curve) of the
secondary; top right: the orbital period; bottom left:
the mass of the secondary (solid curve) and the primary (dashed
curve) and the mass of the hydrogen exhausted core (dotted curve);
bottom left: the mass-loss rate of the secondary.
The evolutionary track of the secondary in the H-R diagram is shown
in Figure~4.}
\label{fig:6}
\end{minipage}
\hfill
\begin{minipage}[b]{0.48\linewidth}
\centerline{\epsfig{file=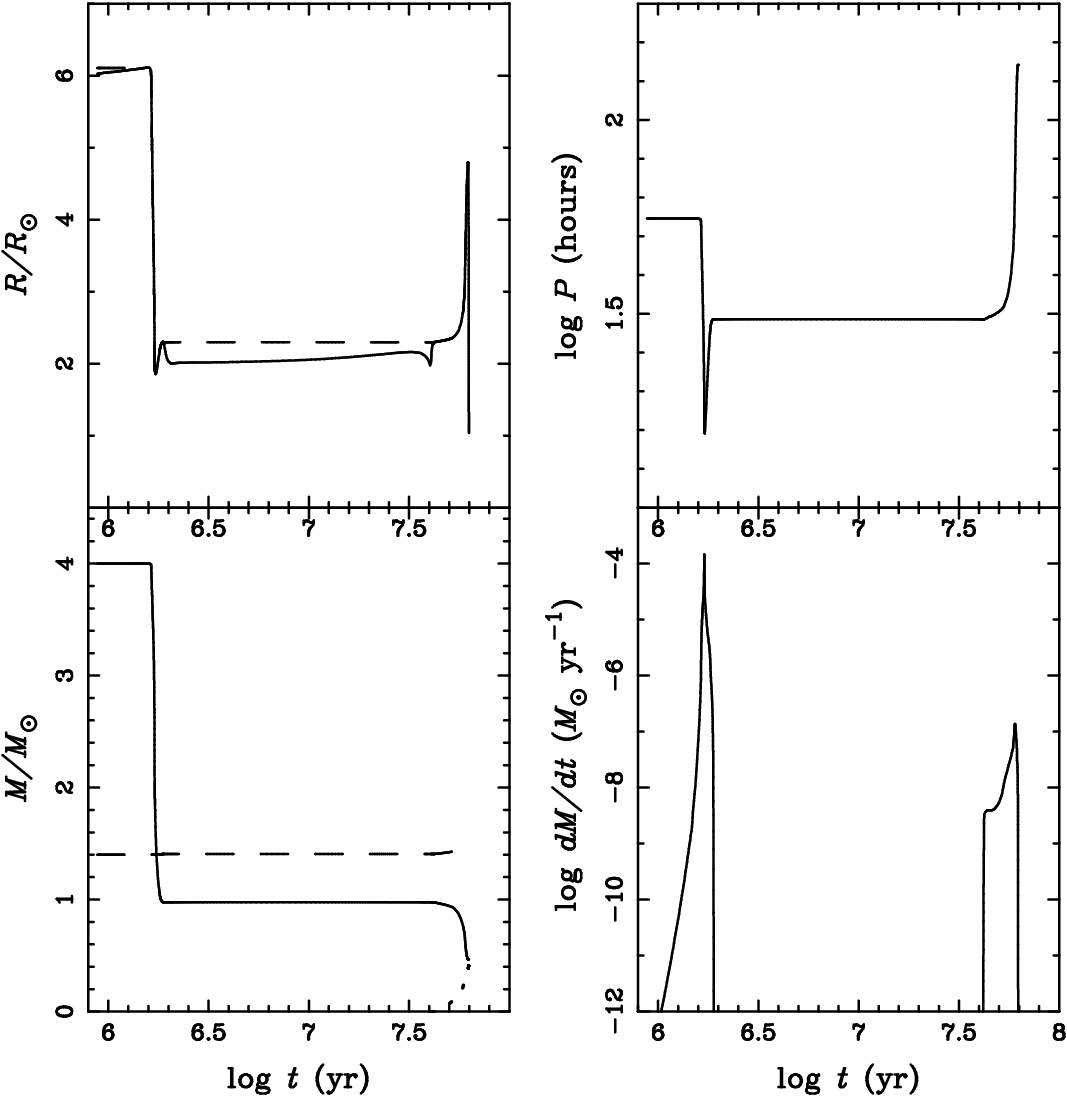,height=\linewidth}}
\caption{Key binary parameters (similar to Fig.~6) for a binary sequence
where the secondary has an initial mass of 4\Ms\ and fills its Roche lobe near
the end of the main sequence (when $X_c=0.1$). The evolutionary
track of the secondary in the H-R diagram is shown in Figure~4. \vspace{15mm}}
\label{fig:fig7}
\end{minipage}
\end{figure*}

Figure~7 shows a more extreme example of thermal timescale mass
transfer where the secondary has an initial mass of 4.0\Ms\ and is
in a similar evolutionary phase as the secondary in Figure~6.
This system is, in fact, on the brink of experiencing a delayed
dynamical instability (see \S~3.3). In this case, equation~(11)
provides a good estimate for the {\em average} mass-transfer rate
in the thermal mass-transfer phase of $\sim 4\times 10^{-6}\Msyr$,
but, as discussed by Langer et al.\ (2000), it does not describe
the detailed behavior of this phase very well. In the turn-on phase, which
lasts of order a Kelvin-Helmholtz time, $\dot{M}$ is significantly
less than $\dot{M}_{\rm th}$, simply because there is so little mass
in the outer layers of the donor star and very little mass needs to
be transferred for the secondary to adjust its radius to the shrinking
Roche-lobe radius (which again is reflected in the large
adiabatic mass--radius exponent). Equation~(11) also does not provide
a good estimate for the maximum $\dot{M}$ of $1.7\times 10^{-4}\Msyr$.
The reason is that, at this high mass-transfer rate,
only the outer layers will be able to adjust thermally and drive
the expansion and that, in this case, $t_{\rm KH}$ in equation~(11)
should be replaced by the shorter thermal timescale of this layer.
What fraction of the envelope can adjust thermally also depends
to a large degree on how the
Roche-lobe radius changes (through the Roche-lobe filling
constraint, i.e., eq.~2), which in turn depends on external factors
such as the change of the mass ratio and systemic mass and angular
momentum loss and not on the internal properties of the secondary.
In the somewhat  extreme example shown in Figure~7, the
secondary evolves essentially adiabatically near the peak in $\dot{M}$.

The secondary will only be able to re-establish thermal equilibrium
once the Roche-lobe radius starts to expand (generally after the mass
ratio has been reversed). At this stage, the secondary will be significantly
undersized and underluminous for its mass. However, this equilibration
phase itself will take a full Kelvin-Helmholtz time and a significant
amount of mass ($\sim 0.4\Ms$ in the sequence shown in Fig.~7) will still be
transferred before the secondary has re-established thermal
equilibrium.

To illustrate this thermal relaxation phase further, we calculated
a separate mass-loss sequence for an unevolved 2.1\Ms\ star losing
mass at a constant rate of $10^{-6}\Msyr$ ($\sim 5 \times \dot{M}_{\rm th}$)
until its mass had been reduced to 1\Ms\ and then let it relax until
it reestablished thermal equilibrium. Figure~8 shows the evolution of the
radius (solid curve) and Figure~9 the evolution of the entropy profile in this
calculation. First note in Figure~9 that only the outer layers of the secondary
(in mass) are able to thermally adjust significantly (and only at early times).
During the mass-loss phase, the radius of the star is always substantially
smaller than the equilibrium radius of a star of the same mass
(shown as a dashed curve in Fig.~8). In the subsequent relaxation phase,
however, the radius overshoots the equilibrium radius by about 15\,\%.
The reason is that a star does not relax in a uniform, homologous
way, but different parts of the star adjust on their local thermal
timescales which vary throughout the star. This mismatch of
timescales drives a thermal wave through the
star (associated with a luminosity wave) which causes the overshooting
in radius (and luminosity). This is also the reason why the system in Figure~7
becomes detached immediately after the thermal mass-transfer
phase\footnote{To the best of our knowledge, this effect was first
noted in calculations of Algol systems by R. P. Pennington in his
Ph.D. thesis (Pennington 1986).}.

\begin{figure*}[t]
\begin{minipage}[b]{0.48\linewidth}
\centerline{\epsfig{file=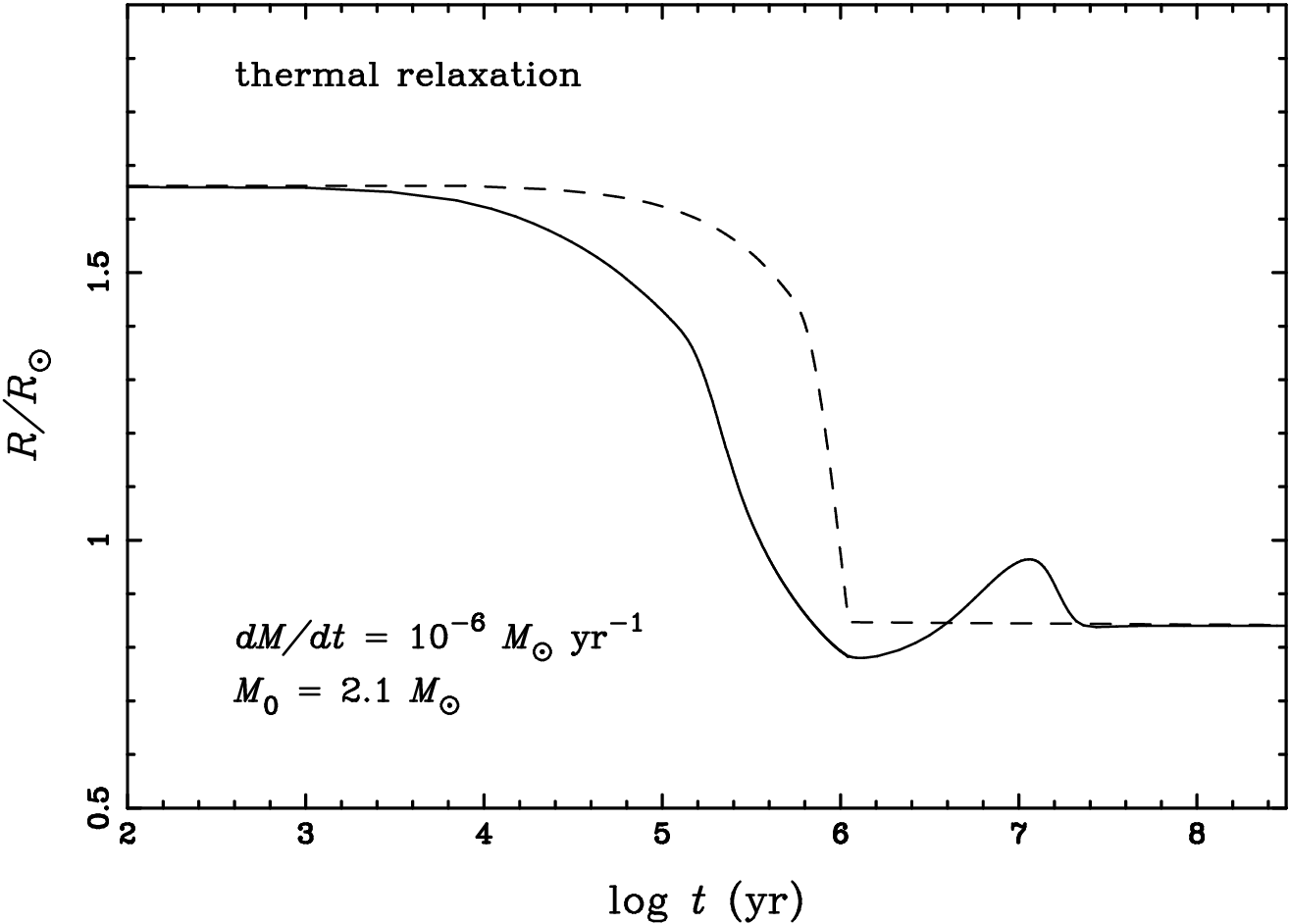,height=\linewidth,angle=-90}}
\caption{Evolution of radius (solid curve) as a function of time for
an unevolved 2.1\Ms\ star losing mass at a constant rate
of $10^{-6}\Msyr$ for $1.1\times 10^6\,$yr until its mass is reduced to
1\Ms.  The dashed curve shows the thermal equilibrium radius of a
star of the same mass at each time.}
\label{fig:8}
\end{minipage}
\hfill
\begin{minipage}[b]{0.48\linewidth}
\centerline{\epsfig{file=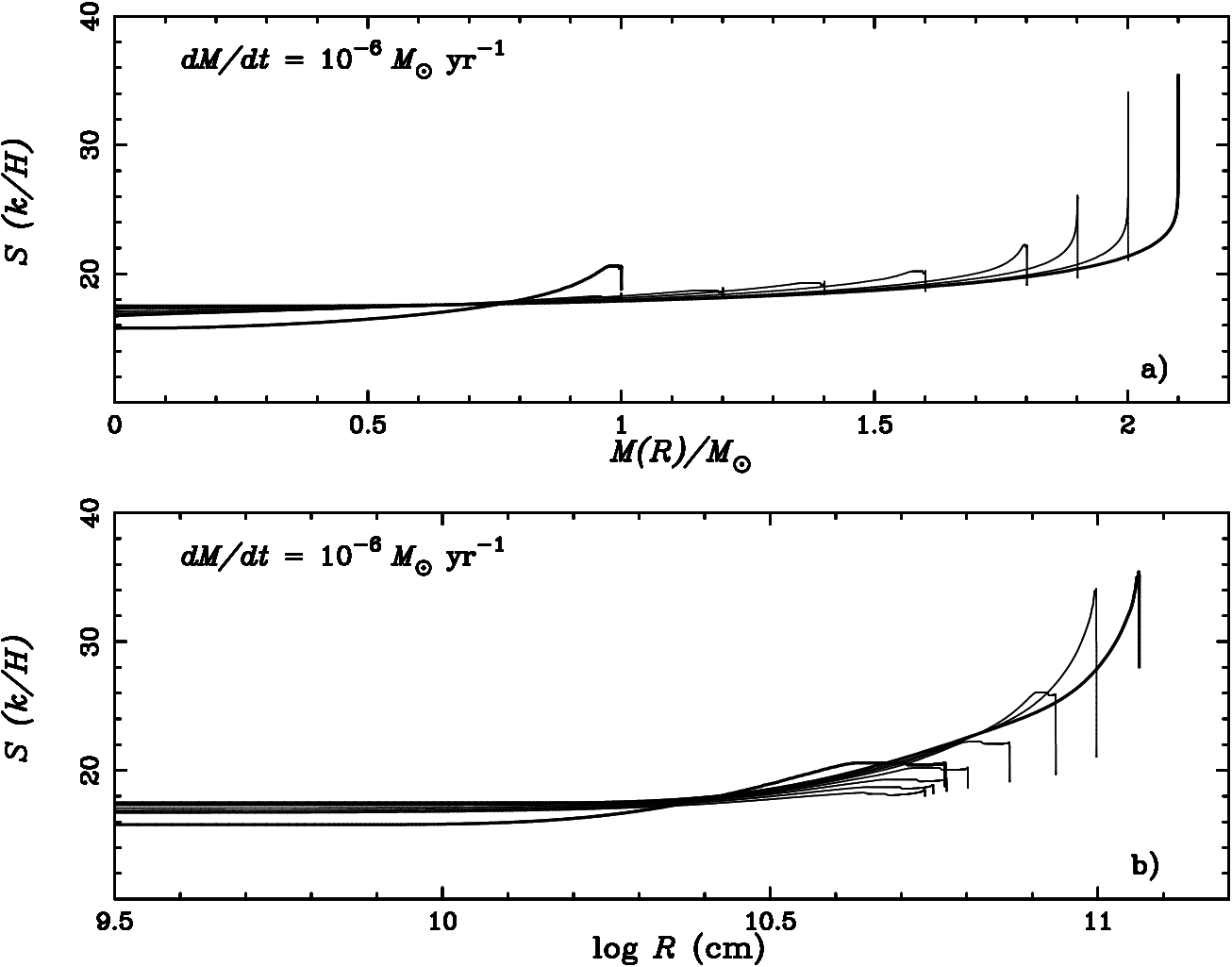,height=\linewidth,angle=-90}}
\caption{Evolution of entropy profile as a function
of mass (top panel) and radius (bottom panel) for the mass-loss and
thermal relaxation calculation shown in Figure~8.
The two thicker curves represent the initial and the final models
in thermal equilibrium. \vspace{2.5mm}}
\label{fig:fig9}
\end{minipage}
\end{figure*}

The binary sequence shown in Figure~6 may, at early times, represent
the evolution for a system like HZ Her/Her X-1, which has an orbital
period of 41\,hr and contains a slightly evolved secondary of $\sim
2.35\Ms$ (e.g., \markcite{Joss}Joss \& Rappaport 1984). At late times,
the sequence may be appropriate for a system like the LMXB X-ray pulsar GRO
J1744-28 with an orbital period of 11.8\,d (\markcite{Finger}Finger
et al.\ 1996) and a donor mass that is likely in the range
$0.2\,$--$\,0.4\,M_\odot$ (\markcite{Rappaport6}Rappaport \& Joss
1997). After the initial high-$\dot{M}$ phase, mass transfer
is driven by the nuclear evolution of the star and starts to rise
towards the end of the main-sequence phase. The system becomes briefly
detached at the point of hydrogen exhaustion (associated with a brief
shrinkage in radius). The subsequent peak in $\dot{M}$ occurs when
magnetic braking has become most active. As the system expands,
magnetic braking becomes less effective and consequently $\dot{M}$
starts to decrease.  At some point nuclear evolution on the sub-giant
branch becomes the dominant mass-transfer driving mechanism, and
$\dot{M}$ rises as the nuclear timescale becomes shorter. Eventually,
the system becomes detached when the secondary has a mass of
0.322\Ms. Despite this low mass, the secondary still ignites helium in
its core and ultimately ends its evolution as a HeCO white dwarf (see
\S 4.1).

Figure~10 shows a sequence similar to Figure~6 for a secondary
with an initial mass of 2.1\Ms, except that it is less
evolved initially (its initial, central hydrogen abundance was
$X_c = 0.489$). This sequence may provide a model for the X-ray
binary Sco X-1 with an orbital period of 18.9\,hr. In this case, the
high observed luminosity of Sco X-1 would be the result of thermal timescale
mass transfer. In this particular model, the secondary of Sco X-1 is
predicted to have a mass of $\sim 2\Ms$ and resemble an A or F star
(absent any X-ray heating effects; see the corresponding
evolutionary track in Fig.~4). After the thermal timescale phase,
mass transfer is driven by the nuclear evolution of the core. As the star
develops a convective envelope, magnetic braking takes over as the dominant
mass-transfer driving mechanism, causing a spike in the mass-transfer rate.
As the system expands, magnetic braking becomes less effective
and the system becomes briefly detached as it evolves up the giant branch.
Eventually, after the secondary has lost most of its hydrogen-rich
envelope, it evolves away from the giant branch and
ends its evolution as a He white dwarf with a mass of 0.231\Ms.

The binary sequence in Figure~7 represents a slightly more massive
version of the evolution that may explain the evolutionary history of
Cyg X-2 (see Podsiadlowski \& Rappaport 2000). After the thermal timescale
phase, the system becomes detached and stays detached for the next
$\sim 4\times 10^7\,$yr. In this phase, the secondary has the
appearance of a $\sim 2.5 \Ms$ main-sequence star in the H-R diagram
(see Fig.~4), except that it is significantly undermassive with a mass
of only $\sim 1\Ms$ and has a low surface hydrogen abundance of 0.34
(by mass). The companion may appear as a slightly spun-up radio pulsar,
having accreted $\sim 0.007\Ms$ of material. The secondary starts
to transfer mass again shortly after exhausting hydrogen in its core.
In this second mass-transfer phase, the evolution is driven by
nuclear shell burning (similar to the case AB model of
Podsiadlowski et al.\ 2000). The final mass of the HeCO white
dwarf is 0.466\Ms.

\begin{figure*}[t]
\begin{minipage}[b]{0.48\linewidth}
\centerline{\epsfig{file=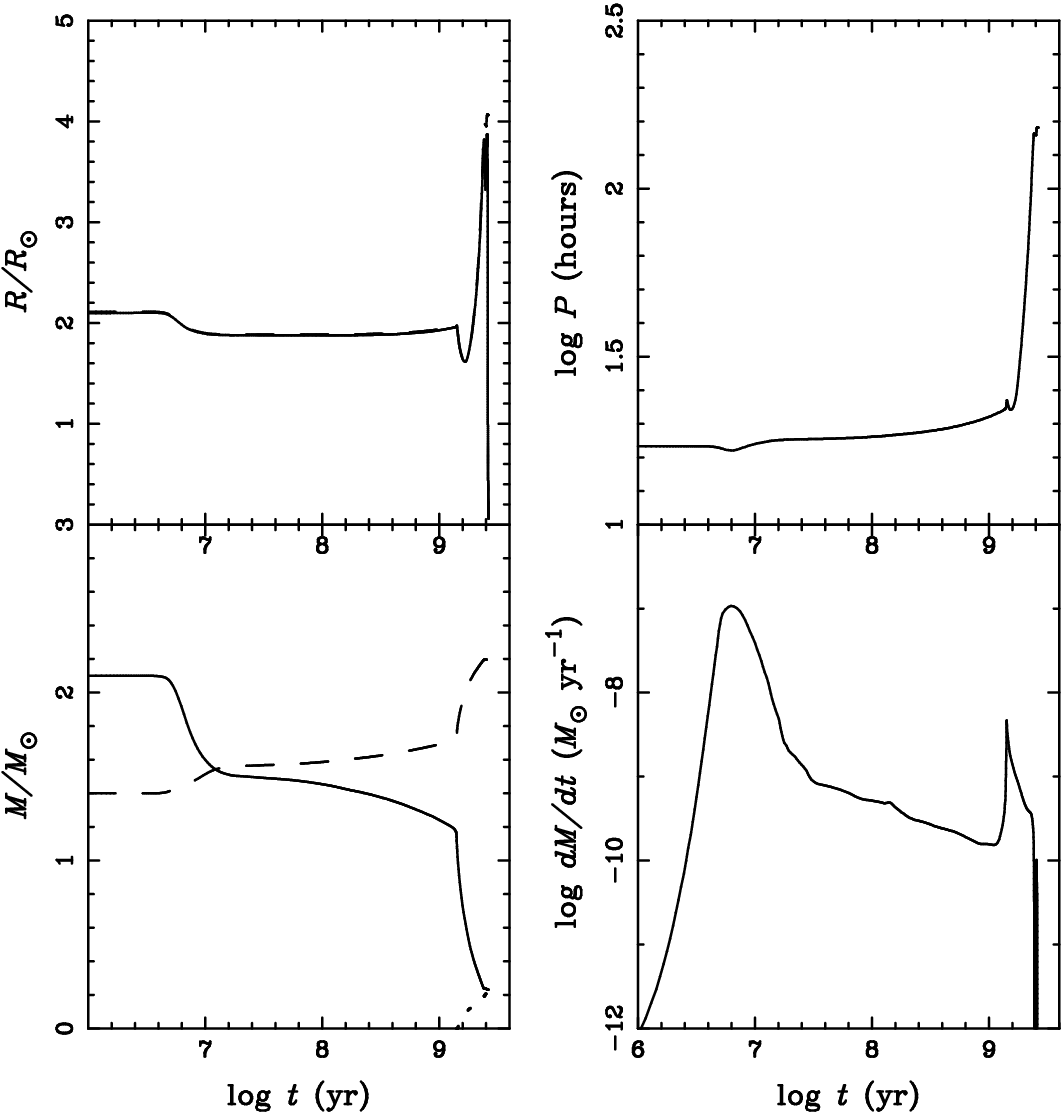,height=\linewidth}}
\caption{Key binary parameters (similar to Fig.~6) for a binary sequence
where the secondary has an initial mass of 2.1\Ms\ and fills its Roche lobe
on the main sequence (when $X_c = 0.489$). The evolutionary
track of the secondary in the H-R diagram is shown in Figure~4.}
\label{fig:10}
\end{minipage}
\hfill
\begin{minipage}[b]{0.48\linewidth}
\centerline{\epsfig{file=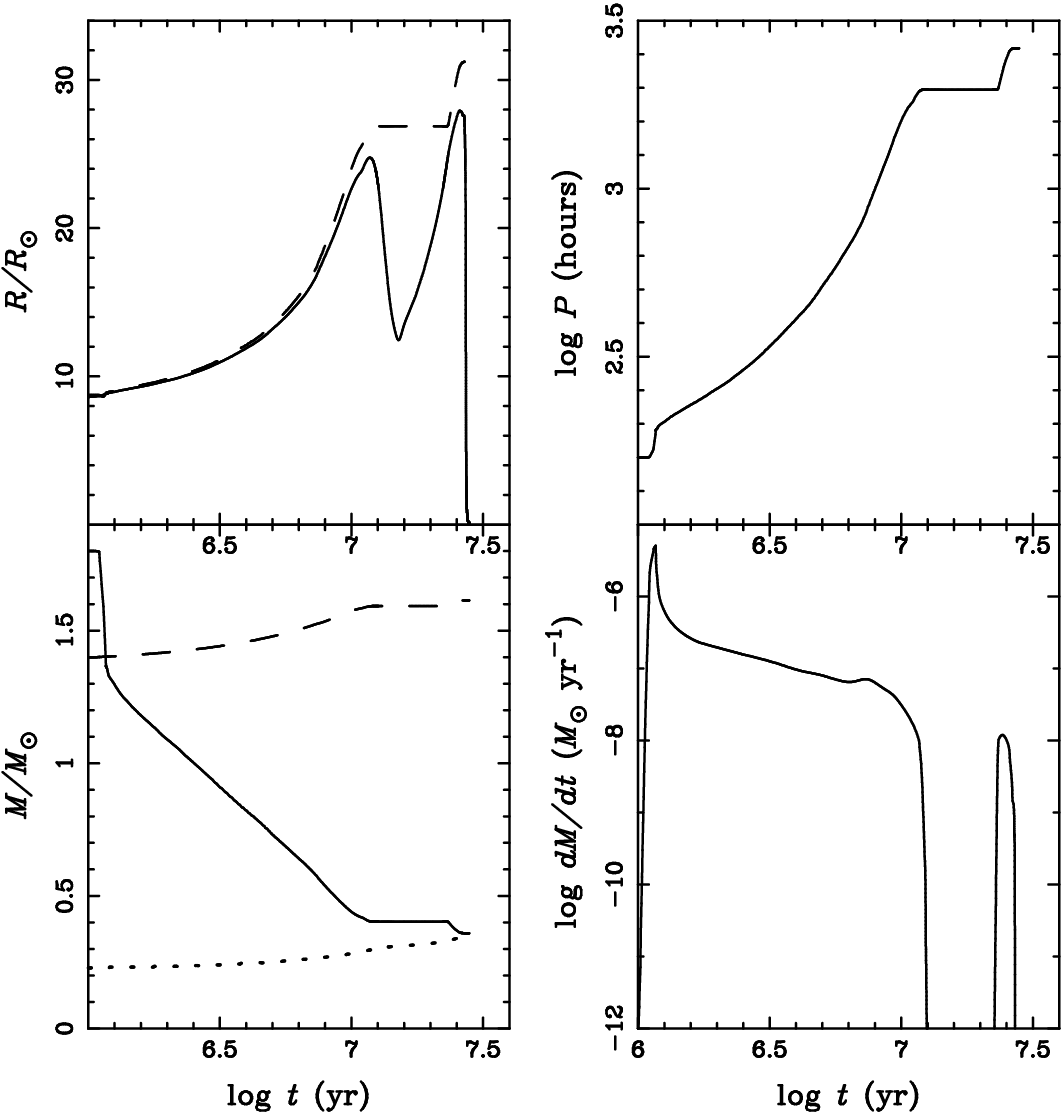,height=\linewidth}}
\caption{Key binary parameters (similar to Fig.~6) for a binary sequence
where the secondary has an initial mass of 1.8\Ms\ and fills its Roche lobe
on the giant branch (when it has a degenerate core of 0.23\Ms).
The evolutionary track of the secondary in the H-R diagram is shown
in Figure~4.}
\label{fig:fig11}
\end{minipage}
\end{figure*}

\subsection{Dynamically unstable mass transfer}
A dynamical mass-transfer instability occurs when the Roche-lobe
radius shrinks more rapidly (or expands less slowly) than the star can
adjust either thermally or adiabatically, i.e.,  when $\xi_{\rm RL} >
\xi_{\rm ad}$ (see \S~3.2). This will then most likely lead to a
common-envelope and a spiral-in phase \markcite{Paczynski3}(Paczy\'nski
1976). For a binary initially consisting of a (sub-)giant and a
neutron star, the system will either merge completely to form a
rapidly rotating single object (a Thorne-\.Zytkow object?
\markcite{Thorne}Thorne \& \.Zytkow 1977) or become a short-period binary
with a white-dwarf companion if the envelope is ejected.

For a fully convective star (approximated by an $n=1.5$ polytrope),
the condition
$\xi_{\rm RL} = \xi_{\rm ad}$ defines a critical mass ratio $q_{\rm
crit}\simeq 3/2$ (where $q=M_1/M_2$; see, e.g., \markcite{Faulkner}Faulkner
1971; \markcite{Paczynski4}Paczy\'nski \& Sienkiewicz 1972;
\markcite{Rappaport3}Rappaport et al.\ 1982).  If the
accreting star is a neutron star of 1.4\Ms, this criterion applied
literally would imply that mass transfer would be dynamically unstable
if the secondary is a \mbox{(sub-)}giant larger than $\sim 0.9\Ms$.
However, (sub-)giants are generally not well
represented by fully convective polytropes. For example,
\markcite{Hjellming}Hjellming \& Webbink (1987) (also see
\markcite{Soberman}Soberman, Phinney, \& van den Heuvel 1997) showed
that the fact that
(sub-)giants have degenerate cores of finite mass can increase $q_{\rm
crit}$ significantly. This criterion also does not take into account
any time delay between the onset of mass transfer and the appearance
of the dynamical instability, during which a substantial amount of mass
may already be transferred in a stable manner.

In our calculations, dynamical instability manifests itself by the
fact that we can no longer satisfy the Roche-lobe filling constraint
in equation~(2). The secondary will subsequently overfill its Roche
lobe by an ever increasing amount. Since we generally limit
the maximum mass-transfer rate to $10^{-4}\Msyr$, some stars near the brink of
dynamical instability will not be able to satisfy equation~(2) in our
calculations,  but would do so without the constraint of a maximum
mass-transfer rate. We therefore assume that all systems where the secondary
overfills its Roche lobe by at most a few per cent for a short amount of time
are stable against dynamical mass transfer\footnote{In actual
fact, this situation arose only in three sequences. In one case,
we recalculated the sequence without the maximum constraint (the
sequence shown in Fig.~7) and, as expected, the system was then able
to fulfill equation~(2) at all times, confirming that it was stable
against dynamical mass transfer.}.
In all other cases, we continued the calculations until the overflow
factor ($f_{\rm over}\equiv R/R_{\rm L}$) exceeded a value of 1.5.

\markcite{Tauris1}Tauris \& Savonije (1999) have recently examined
the dynamical stability for X-ray binaries with \mbox{(sub-)}\-giant donors in
detail, using realistic binary stellar evolution calculations, and
found that all systems with (sub-)giant donor masses $\la 2\Ms$ were
dynamically stable (they assumed an initial neutron-star mass of 1.3\Ms).  In
our calculations we also find that all sequences with donor stars up
to 1.8\Ms\ are dynamically stable, irrespective of evolutionary
phase. Indeed, even for more massive secondaries, we often find that
mass transfer is either dynamically stable (if the secondaries start
mass transfer at the beginning of their ascent of the giant branch) or that
the secondaries overfill their Roche lobes by only a relatively moderate
amount (the most evolved secondaries in our sequences with initial
masses of 2.1, 2.4, and 2.7\Ms\ overfill their Roche lobes
by at most 9, 12, and 13\,\%, respectively; see Table~A1). While
the latter is likely to lead to the formation of a common envelope, it
is not obvious that it necessarily leads to a spiral-in phase, since
there is no friction between the immersed binary and the envelope,
as long as the envelope can remain tidally locked to the orbiting
binary (see, e.g., \markcite{Sawada}Sawada et al.\ 1984).

In Figure~11 we present the binary sequence for the most evolved
1.8\Ms\ secondary we calculated. The initial peak in the mass-transfer
rate is very high ($\sim 9\times 10^{-6}\Ms$), but mass-transfer
remains stable. Even after the mass ratio has been reversed, $\dot{M}$
is significantly super-Eddington. The system becomes temporarily
detached when the H-burning shell starts to move into the region
with a gradient in hydrogen abundance, established during the
hydrogen core burning phase, and the giant shrinks significantly
(\markcite{Thomas}Thomas 1967). We find these temporarily detached
phases in most of our sequences where the secondary evolves up the giant
branch. (These detached phases on the giant branch have also been
found in a number of other recent studies; \markcite{Tauris1} Tauris
\& Savonije 1999; \markcite{Han}Han et al.\ 2000; N. Langer 1999 [private
communication].)

Many of the systems in which the initial secondary mass is
$4\,M_{\odot}$ and probably all systems more massive than
$\sim 4.5\,M_{\odot}$ experience dynamical mass transfer (see Fig.~1
and Table~A1), resulting in the spiral-in of
the neutron star inside the secondary. However, in all systems where
the 
\begin{inlinefigure}\label{fig:fig12}
\centerline{\epsfig{file=fig12.ps,width=0.75\linewidth}}
\caption{Orbital period, radius, and mass-transfer rate (top to bottom) as a
function of time since the beginning of mass transfer (with arbitrary
offset) for an initially unevolved secondary of 4.5\Ms\ to illustrate
the case of a delayed dynamical instability. Phases of
`atmospheric' and `radiative' mass transfer are separated by
a vertical dashed line. The evolutionary track of the secondary in the
H-R diagram is shown in Figure~4.}
\end{inlinefigure}

\noindent secondary is still on the main sequence when mass transfer starts,
this dynamical instability is delayed (see Hjellming \& Webbink 1987)
since the secondaries initially have radiative envelopes with
large adiabatic mass--radius exponents (as discussed in \S~3.2), which
stabilizes them against dynamical mass transfer.
Dynamical instability occurs once the radiative part of the envelope
with a steeply rising entropy profile (the entropy spike near the surface
in Fig.~9) has been lost and the core with a relatively
flat entropy profile starts to determine the reaction of the star to mass
loss. This delay may last for up to $\sim 10^6\,$yr; during this time the
system
should still be detectable as an X-ray binary, with a very high mass-transfer
rate and quite possibly some unusual properties (such as SS433?) in the last
$10^4\,$--$\,10^5\,$yr before the onset of the dynamical instability.

Figure~12 illustrates the case of a delayed dynamical instability
for an initially unevolved 4.5\Ms\ secondary. The early mass-transfer
phase can be divided into two separate phases: (1) a phase of atmospheric
Roche-lobe overflow where $\dot{M}$ increases exponentially
(according to eq.~2) because the radius of the star approaches
the Roche-lobe radius; and (2) a phase (labelled `radiative'
in Fig.~12) where the high-entropy material in the low-density envelope
of the secondary is lost. The binary parameters remain essentially
unchanged in the first phase, lasting $\sim 1.2\times 10^6\,$yr in
this example, but the secondary loses $\sim 0.5\,\Ms$ in the second
much shorter phase, lasting only $\sim 10^5\,$yr, and both the radius
and the orbital period shrink drastically. At the onset of the
dynamical instability (which we here take as the point when
$\dot{M}$ exceeds $10^{-4}\Msyr$), the secondary is extremely
underluminous and has the appearance of a $\sim 1.3\Ms$ main-sequence
star in the H-R diagram (see Fig.~4).

\begin{figure*}[t]\label{fig:fig13}
\centerline{\epsfig{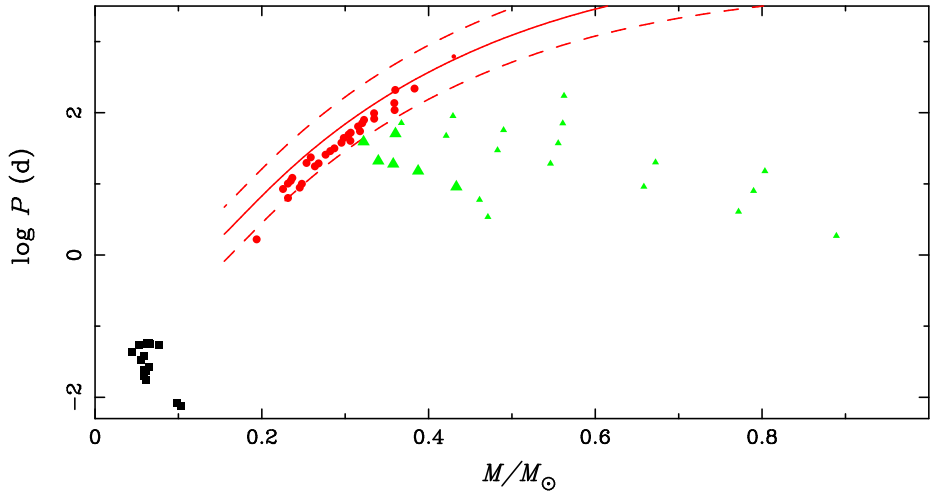}}
\caption{The final distribution of the calculated sequences in the
secondary mass -- orbital period plane (for sequences that
avoided dynamical mass transfer). Circles and
triangles indicate that the secondaries are He white dwarfs and HeCO
white dwarfs, respectively. The size of the symbols indicates how much
mass the neutron stars have accreted (systems with large symbols may
be reasonably expected to contain millisecond pulsars).  The low-mass,
ultracompact systems ($M_{2} < 0.15\,M_{\odot}$, $P_{\rm orb} <
0.1\,$d) are plotted when the systems pass through the orbital-period
minimum.  The solid and dashed curves give the average and the range of
the white-dwarf mass -- orbital-period relation for wide binary
radio pulsars (from \markcite{Rappaport1}Rappaport et al.\ 1995).}
\end{figure*}

\section{End Products}

\subsection{Pulsars with He, HeCO White Dwarfs}

In Figure~13 we show the final distribution of the calculated systems
in the secondary mass -- orbital period plane (for sequences
that avoided dynamical instability). Here
the size of the symbols indicates how much mass a neutron star
has accreted. In systems with large symbols, the neutron star has accreted
at least 0.2\Ms\ (for our accretion prescription) and may be reasonably
expected
to appear as a millisecond pulsar. A circle indicates that the secondary
ends its evolution as a He white dwarf. Note that the He white dwarfs
form a sequence that quite closely follows the relation between white-dwarf
mass and orbital period for wide binary radio pulsars as calculated by
\markcite{Rappaport1}Rappaport at al.\ (1995; solid and dashed curves).
The new sequence may be somewhat steeper at low masses (also see
\markcite{Ergma1}Ergma 1996;
\markcite{Tauris1}Tauris \& Savonije 1999) and lies systematically
below the average sequence of \markcite{Rappaport1}Rappaport et al.\
(1995). The latter can be easily understood since, when the secondaries
become detached from their Roche lobes, they have already evolved
somewhat away from a Hayashi track and are hotter (and hence smaller)
than a giant of the same core mass (see Fig.~4), an effect that could
not be easily included without detailed binary evolution calculations. This
suggests that one should rescale the average relation of \markcite{Rappaport1}
Rappaport et al.\ (1995) by a factor $\sim 0.65$.

In systems with triangle symbols, the secondaries ignite helium in the
core and generally burn helium in a hot OB subdwarf
phase (after mass transfer has been completed). Note that the lowest
mass of a helium star for which helium can be ignited is $\sim 0.3\Ms$,
the minimum mass for helium ignition in non-degenerate
cores (see, e.g., \markcite{Kippenhahn3}Kippenhahn \& Weigert 1990).
While the more massive
helium stars convert most of their mass into carbon and oxygen
(typically having a helium-rich envelope of at most a few per cent),
the lower-mass helium stars only burn helium completely in the core
and end their evolution with large helium envelopes (this was
found first in calculations by \markcite{Iben}Iben \& Tutukov [1985] and
more recently by\markcite{Han} Han et al.\ [2000]). It is not clear at
the present time whether the fact
that these low-mass HeCO white dwarfs have large CO cores
has detectable, observational consequences. The most interesting
aspect of the systems with HeCO white dwarfs is, of course,
that most of them lie well below  the white-dwarf -- orbital period
relation  without having experienced a  common-envelope phase
\begin{inlinefigure}\label{fig:fig14}
\centerline{\epsfig{file=fig14.ps,width=\linewidth}}
\caption{Hydrogen shell flashes for a 0.199\Ms\ helium white dwarf
originating from a secondary with an initial mass of 1.4\Ms:
H-R diagram (panel a); evolution of luminosity and radius as a function
of time since the beginning of the initial mass-transfer phase for
all flashes (panels b and d) and for just the third flash (panel
c and e) with arbitrary time offset.}
\end{inlinefigure}

\vspace{1cm}

\noindent (also see \markcite{Podsi1}Podsiadlowski \& Rappaport 2000;
\markcite{Tauris2}Tauris et al.\ 2000).

Finally, it is worth noting that most of the He white dwarfs with masses
$< 0.4\,M_{\odot}$  and even some of the more massive HeCO white dwarfs
experience several dramatic hydrogen shell flashes
(typically 2 to 4) before settling on the white-dwarf cooling sequence.
These flashes have been extensively discussed in the literature
(e.g., \markcite{Kippenhahn3}Kippenhahn, Thomas, \& Weigert 1968;
\markcite{Iben2}Iben \& Tutukov 1986). More
recently, \markcite{Sarna}Sarna, Ergma, \&  Ger\v skevit\v s-Antipova
(2000) published a detailed study of hydrogen shell flashes for low-mass He
white dwarfs and their implications for the calculations of cooling
ages in companions of binary millisecond pulsars.

During these flashes, the luminosity typically rises by a factor of
1000 and the radius increases by a factor of 10 or more on timescales
of a few decades. Indeed, during these flashes the secondaries tend to
fill their Roche lobes again, leading to several short mass-transfer
phases with mass-transfer rates that are often much higher than
the rates achieved in earlier phases (typically with
$\dot{M}\sim \mbox{several}\times 10^{-6}\Msyr$). In Figure~14 we present an
example of a He white dwarf of 0.199\Ms\ which experiences 3 such flashes
(the secondary originally had a mass of 1.4\Ms\ and filled
its Roche lobe near the end of the main sequence).

\begin{figure*}[t]
\centerline{\epsfig{file=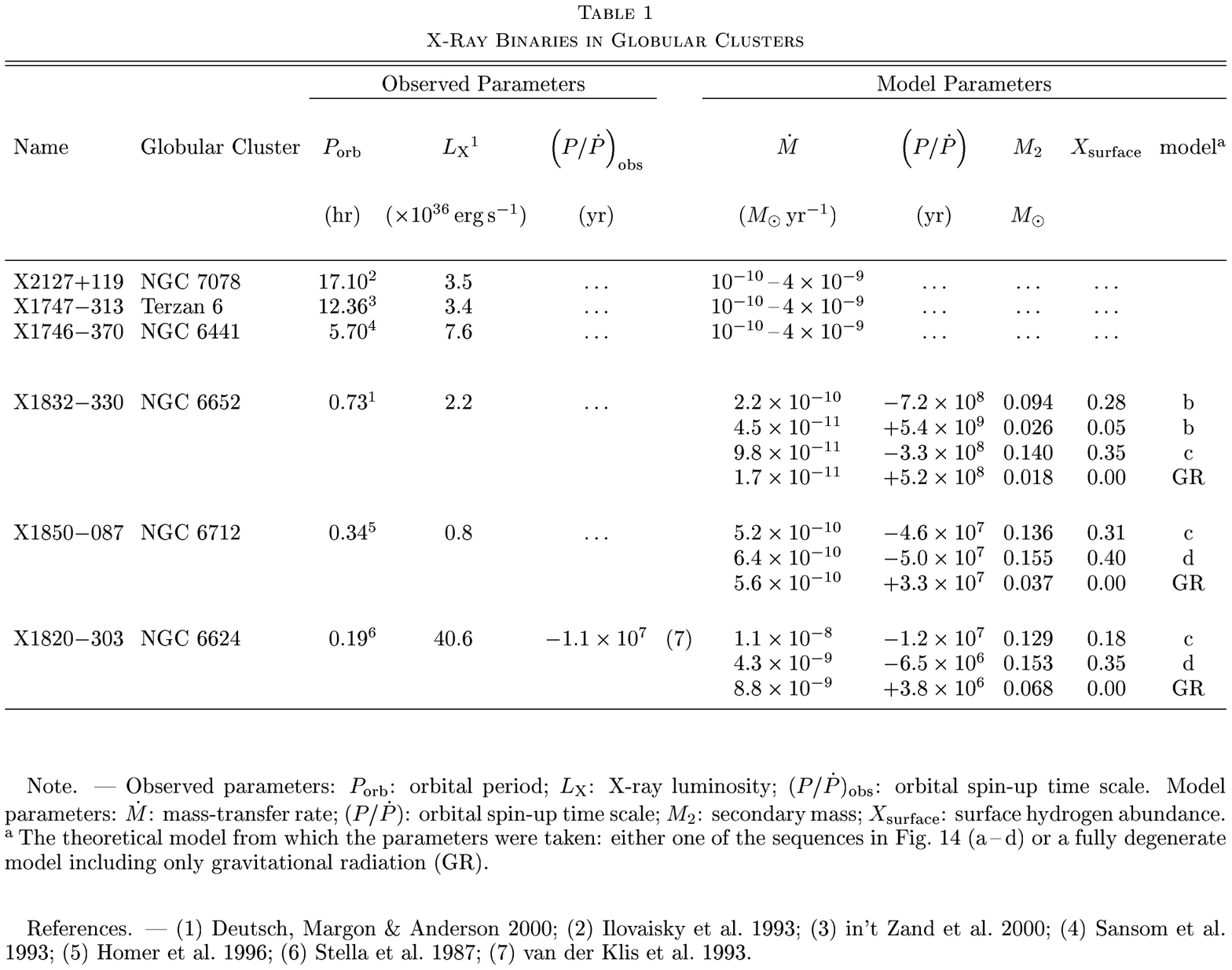,width=0.8\linewidth}}
\end{figure*}

\subsection{Ultracompact X-ray Binaries}

As Figures~2a and 13 show, systems with initial orbital periods below
the bifurcation period ($\sim 18\,$hr) become ultracompact binaries
with minimum orbital periods in the range of 11$\,$--$\,83\,$min. The
shortest period is similar to the $11\,$min period in the X-ray binary
4U 1820-30 in the globular cluster NGC 6624 (\markcite{Stella}Stella
et al.\ 1987). Unlike the two better-known models for the formation of
this system, this evolutionary channel involves neither a direct
collision (\markcite{Verbunt2}Verbunt 1987) nor a common-envelope
phase (\markcite{Bailyn}Bailyn \& Grindlay 1987;
\markcite{Rasio}Rasio, Pfahl, \& Rappaport 2000)  and therefore
constitutes an attractive alternative scenario for 4U 1820-30. This
alternative
evolutionary path for the origin of 4U1820-30 was originally suggested
by \markcite{Tutukov2}Tutukov et al.\ (1987)\footnote{Another
model suggested for 4U 1820-30 involves a non-degenerate, helium-star
companion (Savonije, de Kool, \& van den Heuvel 1986). However, such a model
requires a  complex triple-star interaction in order to form a non-degenerate
helium star at the present epoch (see the discussion in van der
Klis et al.\  1993).}.
\markcite{Fedorova}Fedorova \& Ergma (1989) made the first
detailed case study of this scenario and showed that, if mass transfer
starts near or just after the point of central hydrogen exhaustion,
orbital periods as short as 8 min could be attained and that a system
like 4U 1820-30 can pass through an orbital of 11 min twice, while
approaching the period minimum and after having passed through it. In
related studies,
\markcite{Nelson}Nelson, Rappaport \& Joss (1986) and
\markcite{Pylyser1}Pylyser \& Savonije (1988)
found minimum periods as short as 34 and 38 min, respectively. These
authors were not specifically trying to explain 4U 1820-30 but
their Galactic counterparts 4U 1626-67 and 4U 1915-05 with
orbital periods of 41\,min (\markcite{Middleditch}Middleditch et al.\ 1981;
\markcite{Chakrabarty}Chakrabarty 1998) and 50\,min
(\markcite{Chou}Chou, Grindlay, \& Bloser 2001), respectively. \par

\begin{figure*}[t]
\begin{minipage}[b]{0.48\linewidth}
\centerline{\epsfig{file=fig15.ps,width=0.95\linewidth}}
\caption{Formation of ultracompact LMXBs. From top to bottom:
the minimum orbital period, the mass of the secondary, the mass-transfer
rate, and the surface hydrogen abundance (all at the minimum period) versus
initial orbital period. All calculations start
with a $1.4\,M_{\odot}$ neutron star and a $1\,M_{\odot}$ secondary.
The vertical dashed line indicates the initial orbital period above which
systems become wider rather than more compact (i.e., the bifurcation period).
The dots indicate the results of the calculated sequences. The dashed
curves in the top panel indicate the range of minimum periods if mass
transfer is assumed to be either fully conservative (upper curve) or fully
non-conservative (lower curve).}
\label{fig:15}
\end{minipage}
\hfill
\begin{minipage}[b]{0.48\linewidth}
\centerline{\epsfig{file=fig16_c.ps,height=0.965\linewidth,angle=-90}}
\caption{Selected binary sequences illustrating the formation of ultracompact
LMXBs: sequence (a; black), sequence (b; green), sequence (c; red),
sequence (d; blue). The individual panels show the evolutionary tracks
in the H-R diagram (top), and the evolution of orbital period (middle)
and mass-transfer rate (bottom) as a function of time since the
beginning of mass transfer.  The open circles show where the systems
in the individual sequences reach their period minimums.
Other symbols indicate when the systems pass through the periods of the
three ultracompact LMXBs in globular clusters presently
known (triangles: X1832-330; squares: X1850-087, stars: X1830-303).}
\label{fig:fig16}
\end{minipage}
\end{figure*}

\begin{figure*}[t]\label{fig:fig17}
\centerline{\epsfig{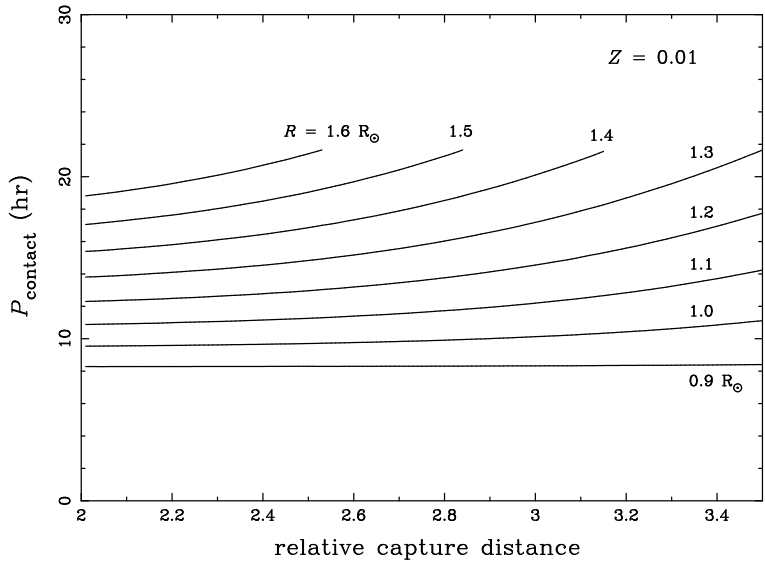}}
\caption{Contact period as a function of relative capture distance for
X-ray binaries that formed as a result of tidal capture. The contact
period defines the initial orbital period at which mass transfer
commences. The relative capture distance gives the ratio of the
initial periastron distance to the radius of the star (with $M_2=1\Ms$
and $Z=0.01$) captured by a neutron star (with $M_1=1.4\Ms$). Different
curves show the contact period for different radii (i.e., evolutionary
stages) of the captured star at the epoch of capture. It is assumed
that the systems are brought into contact by the combined effects
of magnetic braking and evolution of the secondary.}
\end{figure*}

To determine the shortest orbital period that can be attained through this
channel, we performed a separate series of binary calculations for
a $1\,M_{\odot}$ secondary with parameters appropriate for
4U 1820-30 in the metal-rich globular cluster NGC 6624 (i.e.,
with Z=0.01, Y=0.27).  Our results confirm the earlier results
of Fedorova \& Ergma (1989) that, if the
secondaries start mass transfer near the end of core hydrogen burning (or, in
fact, just beyond), the secondaries transform themselves into
degenerate helium stars and that orbital periods as short as
$\sim 5\,$min can be attained without the spiral-in of the neutron star
inside a common envelope. The top panel in Figure~15 shows the relation
between initial orbital period and the minimum period, while the
other panels give the secondary mass, $M_2$, mass-transfer rate, $\dot{M}$,
and surface hydrogen abundance, $X_{\rm s}$, at the minimum
period. There is a fairly large range of initial orbital periods
($13\,$--$\,17.7\,$hr) which leads to ultra-compact LMXBs with a
minimum orbital period of less than $30\,$min. The drop in $P_{\rm min}$
at $13\,$hr
occurs for a model where the secondary has just exhausted hydrogen
in the center at the beginning of mass transfer. The shortest
minimum period is attained for systems just below the bifurcation
period (in this case $\sim 18\,$hr).  The mass-transfer rate at the minimum
period increases significantly as the minimum period decreases,
simply because the time scale for gravitational radiation, which drives
the evolution at this stage, becomes so short.

Figure~16 shows the details of four representative sequences:
sequence (a; black) has an initially unevolved secondary,
in sequence (b; green) the secondary has just completed hydrogen
burning in the center, while in sequences (c; red) and (d; blue)
the secondaries have pure helium cores of 0.024\Ms\ and
0.028\Ms, respectively, at the beginning of mass transfer.
Open circles show when the sequences reach the period minimum,
and other symbols indicate when they pass through the periods
of the three known ultracompact X-ray binaries in globular
clusters (see Table~1). The shortest minimum periods
are obtained for systems that exhaust all of the hydrogen left
in their surface layers just near the point
where they become degenerate and hence manage to transform themselves
into essentially pure helium white dwarfs. In the phase
where the hydrogen shell is being extinguished, the luminosity of the
secondary briefly increases in the sequences with the shortest minimum periods
(top panel), before the secondaries descend on the cooling sequence
for He white dwarfs. In sequence (d; blue), the secondary becomes
detached at an orbital period of 4.3\,hr. Gravitational radiation then
causes the system to shrink, and the secondary starts
to fills its Roche lobe again at an orbital period of 35\,min.
Note also that after the minimum period, $\dot{M}$ generally drops
dramatically.

In Table~1 we list all X-ray binaries in globular clusters with
known orbital periods. Strikingly, three of the six systems
have ultra-short periods.  In this table, we also list the timescales
for the orbital period changes for the three compact systems
based on the sequences shown in Figure~16 as well as on a simple
model where the secondary is a fully degenerate He white dwarf and the
system is
driven by gravitational radiation alone (labelled `GR').
The 11\,min system (4U 1820-30 =  X1820-303) is particularly interesting
since its orbital period is observed to be decreasing rather than increasing
(\markcite{Tan}Tan et al.\ 1991), as would be
expected for a fully degenerate secondary. While it had
been argued that this apparent orbital period decrease could
be caused by gravitational acceleration within the globular
cluster (\markcite{Tan}Tan et al.\ 1991), \markcite{Klis}van der Klis et
al.\ (1993) subsequently concluded that, using a more realistic mass model for
the globular cluster NGC 6624, it was unlikely that the
negative $\dot{P}$ could be fully explained by cluster acceleration.
On the other hand, as Table~1 shows, the theoretical $P/\dot{P}$ in
sequence (c; red) is in excellent agreement with the observed
value, and the value in sequence (d; blue) is in reasonable agreement
(\markcite{Fedorova}Fedorova \& Ergma [1989] obtained similar
values). This provides additional support for this evolutionary
channel for these ultracompact globular-cluster systems.  One
potentially testable prediction is that some of these systems still contain
hydrogen in their envelopes (up to 40\,\% by mass in the sequences
shown). After the period minimum, the secondaries in sequences (c; red) and
(d; blue) become pure He white dwarfs, and their evolution is identical
to that of systems with He white dwarfs, driven by gravitational
radiation alone.

Since the value of the bifurcation period is sensitive to the adopted
magnetic-braking law, the range of initial orbital periods which leads
to ultracompact systems will also depend on it. To examine this
dependence, we carried out two additional series of calculations where
we increased and decreased the efficiency of magnetic braking by a
factor of 5 with respect to our standard model, respectively. As
expected, the bifurcation period increased to $\sim 20\,$hr for the
more efficient magnetic-braking law and decreased to $\sim 15\,$hr for
the less efficient one. In both cases, we obtained systems with
minimum periods as short as 9 and 16\,min, respectively (note,
however, that this exploration was not as comprehensive as for the
standard case).

It is quite remarkable that one of our binary sequences (sequence
c; red), with an initial orbital period around $17\,$hr, appears to be a
suitable sequence to explain all six LMXBs in globular clusters whose
orbital periods are presently known, from the system with the longest
period (AC211/X2127+119 in M15; $P_{\rm orb}=17.1\,$hr; Ilovaisky et
al.\ 1993) to the 11-min binary.  Furthermore, systems with an initial
period in the range of $13\,$--$\,18\,$hr are quite naturally produced
as a result of the tidal capture of a neutron star by a main-sequence
star (\markcite{Fabian}Fabian, Pringle, \& Rees 1975;
\markcite{Distefano2}Di\,\,Stefano \& Rappaport 1992) and are not the generally
expected outcome of a 3- or 4-body exchange interaction (see, e.g.,
\markcite{Rasio}Rasio et al.\ 2000).

To illustrate this, we calculated the orbital period at which mass
transfer commences, the contact period, for systems that form as a
result of the tidal capture of a 1\Ms\ normal star by a 1.4\Ms\
neutron star.  Figure~17 shows the contact period as a function of the
relative capture distance (i.e., the ratio of initial
periastron distance to the radius of the star) for different radii
(i.e., evolutionary stages on the main sequence) of the secondary at the
epoch of capture. In these calculations we assumed that the systems formed by
tidal capture circularized quickly (on a timescale short compared to
the magnetic-braking timescale and the evolutionary timescale of the
secondary) and that the system was brought into contact by the
combined effects of magnetic braking (causing the orbit to shrink) and
the evolution of the secondary.

It has been estimated that for the tidal capture of a main-sequence
star by a neutron star, the initial periastron distance has to be $\la
3$ stellar radii (\markcite{Fabian}Fabian et al.\ 1975;
\markcite{Press2}Press \& Teukolsky 1977; \markcite{McMillan}McMillan,
McDermott, \& Taam 1987). It also has to be larger than $\sim 1.5$
stellar radii, so that in the circularized system (which has a
separation twice the initial periastron distance) the secondary
underfills its Roche lobe. As Figure~17 shows, this range of initial
capture distances produces systems which start mass transfer between 9
and 21\,hr, depending on the radius of the secondary at the time of
capture. The range of radii chosen corresponds to the change in radius
of a 1\Ms\ star on the main sequence (i.e., as it evolves from
the zero-age main sequence to the terminal-age main sequence).
Since all stars on the main
sequence have roughly equal probability for a dynamical encounter with
a neutron star, this predicts a fairly uniform distribution of contact
periods in this range. In fact, stars with a larger radius are
somewhat more likely to be captured since the capture cross section
increases linearly with radius (Fabian et al.\ 1975;
\markcite{Distefano2}Di\,\,Stefano \& Rappaport 1992). This suggests
that at least half of the X-ray binaries formed by tidal capture may
start mass transfer with orbital periods in the range of 13 to 18\,hr,
the range which produces ultracompact systems. This may explain the
surprisingly large fraction of ultracompact systems (3 out of 6 systems
with known orbital periods) in globular clusters.

One well-recognized problem with the tidal-capture scenario for the
formation of LMXBs in globular clusters is the fact that the total
energy that needs to be tidally dissipated during the capture and
subsequent circularization process is of order the binding energy of
the secondary. Since all of this energy is deposited in the secondary,
it may lead to its destruction either by dynamical effects
(\markcite{Rasio2}Rasio \& Shapiro 1991) or due to the thermally
driven expansion of the secondary that is tidally heated
(\markcite{McMillan}McMillan et al.\ 1987; \markcite{Ray}Ray, Kembhavi,
\& Antia 1987;  \markcite{Podsi2}Podsiadlowski 1996). As a consequence, the
initial capture distance for which a 1\Ms\ star can be captured and survive
the process may be much more limited than the range used above. Indeed,
this is the reason why it has been popular in recent years to dismiss
this formation channel altogether. However, we would like to emphasize
(1) that the details of the tidal capture process, in particular the
response of the secondary to tidal heating (see
\markcite{Podsi2}Podsiadlowski 1996), are
still rather uncertain; (2) that, as shown here, tidal capture
naturally produces the range of orbital periods actually observed in
globular-cluster LMXBs; and (3) that alternative scenarios, 3- or
4-body interactions, do not generally lead to systems in the observed
range. This suggests to us that it is not only premature to rule out
tidal capture as a formation scenario for LMXBs, but that the LMXBs in
globular clusters with well-determined orbital periods actually
provide observational evidence in its favor. An unbiased
re-examination of the whole process is therefore clearly warranted.

\begin{figure*}[t]\label{fig:fig18}
\centerline{\epsfig{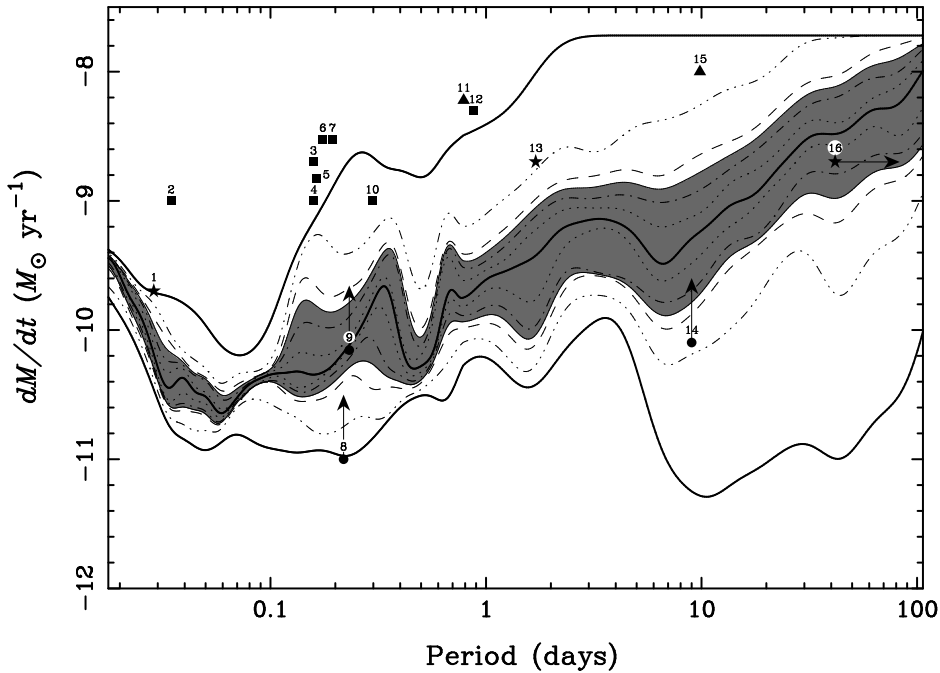}}
\caption{Cumulative (smoothed) probability distribution for the mass-accretion
rate onto the neutron-star primary as a function of orbital period for
the 100 binary calculations (with equal weighting for all
sequences). The thick, solid central curve gives the median
mass-accretion rate; the pairs of curves moving progressively outwards
from the median curve include 20, 40, 60, 80, and 98 per cent of the
distribution, respectively. The shaded region contains 50 per cent of
all systems around the median. The symbols indicate the mass-transfer
rates of selected observed X-ray binaries (triangles: Z sources;
squares: atoll sources; stars: X-ray pulsars; circles: systems with
accretion-disk coronae [lower limits]).  The individual systems are
(in ascending order of orbital period, as given in parentheses): (1)
1626-67 (0.69\,hr), (2) 1916-053 (0.83\,hr), (3) 1636-536 (3.8\,hr),
(4) 0748-676 (3.82\,hr), (5) 1254-690 (3.93\,hr), (6) GX9+9 (4.2\,hr),
(7) 1735-555 (4.65\,hr), (8) 2129+470 (5.24\,hr), (9) 1822-37
(5.57\,hr), (10) 1658-29 (7.11\,hr), (11) Sco X-1 (18.9\,hr), (12),
1624-590 (21\,hr), (13) Her X-1 (40.8\,hr), (14) 0921-630 (216\,hr),
(15) Cyg X-2 (236\,hr), (16) GX1+4 ($> 1000\,$hr) (taken from the
catalogs by \markcite{Paradijs2}van Paradijs 1995 and
\markcite{Swank}Christian \& Swank 1997).}
\end{figure*}

\section{Application to the Population of X-Ray Binaries}

In the previous sections we have presented the results of our 100
binary evolution models for LMXBs and IMXBs which cover a systematic
grid of binary parameters at the onset of mass transfer. We have shown
specifically that a number of these binary sequences pass through
states which would closely resemble many of the well known and
individually studied LMXBs and IMXBs. This includes X-ray binaries
with long and short orbital periods, very low- to moderate-mass
secondaries, a wide range of X-ray luminosities, and systems in and
out of globular clusters. All but a few of our 100 binary evolution
sequences started with donor stars of mass $> 1 M_{\odot}$, and the
majority had donors $> 2 M_{\odot}$.  One of the more striking results
of these calculations is that most of the evolution time these systems
spend as an X-ray binary occurs {\em after} the mass of the donor star
has been reduced to $\la 1 M_\odot$ (see Fig.~2a). Thus, a large
fraction of the systems which we commonly refer to as ``LMXBs'' may
actually have started their lives as ``IMXBs.'' This has important
implications for both (i) the retention of neutron stars in binaries
at the time of the supernova explosion which gives birth to the
neutron star (i.e., it is easier to keep the neutron star bound with a
2 or 3 $M_{\odot}$ companion than with a truly low-mass companion),
and (ii) the evolutionary state of the companion stars in LMXBs that
we observe today (i.e., they are probably not nearly as {\it
un}evolved as was previously assumed).

One observational consequence is that many of these systems should be
hydrogen deficient and helium enriched, and that the surface
composition of many secondaries should show evidence for CNO
processing (i.e., be enhanced in N and depleted in C and O).  Our
calculations predict that at a particular orbital period ($\ga
80\,$min), the surface hydrogen abundance can vary typically between
$\sim 1/3$ of the solar value and solar (at shorter orbital periods,
all systems should be hydrogen-deficient).  While the hydrogen (or
helium) abundance can usually not be measured directly, an increased
helium abundance affects the behavior of accretion, e.g., by
increasing the Eddington accretion rate. This could provide some
indirect evidence for helium enrichment in some of these systems.

While it is, of course, rather gratifying to be able to ``explain''
possible evolutionary paths leading to some of the best known X-ray
binaries, there remain several outstanding issues: (1) how unique are
the evolutionary paths we have found; (2) are types of systems
suggested by other reasonably long-lived phases of our binary
evolutions represented in the observed binary X-ray source population;
and (3) is our complete ensemble of binary evolution models consistent
with the overall population of observed LMXBs and IMXBs?

In order to properly investigate these questions one needs to carry out
a full binary population synthesis (BPS) study, starting from
primordial binaries and utilizing a library of binary evolution
models of the type we have generated.  Such a BPS study is beyond the
scope of the present paper, but has been initiated (\markcite{Pfahl}Pfahl
et al.\ 2001). One objective of a
BPS study will be to produce probability distributions, at the current
epoch, for finding LMXBs and IMXBs with various values of $M$, $\dot
M$, $P_{\rm orb}$, as well as in different evolutionary states,
locations in the Galaxy, space velocities, and so forth. Such a study
will involve weighting each of the binary evolutions in our library
by the probability that each of the initial binary parameters would be
realized in nature. For the present study we utilize our evolution
tracks to produce a simplified estimate of the likelihood of finding
LMXBs and IMXBs in various locations in the $\dot M-P_{\rm orb}$
plane. The choice of these two parameters -- $\dot M$ and $P_{\rm orb}$
-- is motivated by the fact that these are the easiest to determine for
LMXBs. In fact, observationally, very few LMXBs and IMXBs (HZ Her/Her X-1,
Cyg X-2) have well determined constituent masses or have much known
about the state of the donor star. On the other hand, if an estimate
of the distance is known, then $\dot M$ can be inferred from the X-ray
luminosity (at least in the case of conservative mass transfer), and the
orbital period can be inferred from X-ray or optical photometry,
rather than requiring Doppler measurements of either the companion
or the neutron star.

To relate our binary models to the Galactic population of LMXBs and
IMXBs on a statistical basis, we have constructed a plot which
estimates the probability of finding an LMXB or IMXB in a particular
region of the $\dot M-P_{\rm orb}$ plane. We do not attempt to weight
each of the binary evolution runs (in the library). We do, however,
take into account the amount of time spent in a particular part of
the evolution, as each of our model binaries traverses the $\dot
M-P_{\rm orb}$ plane. We make the implicit assumption of a
steady-state production of LMXBs and IMXBs which then proceed through
their entire evolution, well within the lifetime of the Galaxy. This,
of course, will become less valid for systems with evolutionary phases
comparable to the age of the Galaxy. To construct our probability
distribution in the $\dot M-P_{\rm orb}$ plane, we proceeded as
follows. First, we divided up the $\dot M-P_{\rm orb}$ plane into a
finely spaced, discrete, two-dimensional array. Each of the 100 binary
evolution tracks was then placed into this array, weighted by the
evolution time spent in each element of the array. The probability of
finding an LMXB/IMXB in any particular array element is then
proportional to the combined evolution time of all the tracks passing
through that array element. However, since there are only 100
evolution tracks, the entire $\dot M-P_{\rm orb}$ plane is not
completely sampled (for an analogous sampling effect in the $M-P_{\rm
orb}$ plane see Fig. 2a). In order to circumvent this problem somewhat,
we computed, for each value of $P_{\rm orb}$, a cumulative probability
distribution in $\dot M$. We then utilized these to compute contours
of constant probability which are plotted in Figure 18. The central
heavy curve is the median value of $\dot M$, while the contours on
either side are in increments of $10\%$ in probability, except for the
top and bottom curves which represent $1\%$ and $99\%$ of the
systems. The shaded region represents $50\%$ of all systems around
the median.

As one can see from a perusal of Figure 18, there should theoretically
be a general positive correlation between orbital period (for $P_{\rm
orb} \ga 1$ hr) and $\dot M$, with the value of $\dot M$ a few hundred
times larger at periods of $\sim$100 days as compared with 1 hr. At a
given $P_{\rm orb}$, typically half of the systems are contained
within a range of about a factor of $\sim 6$ in $\dot M$, centered on
the median value.  The next step is to compare the model results shown
in Figure 18 with the positions of known LMXBs and IMXBs in this
diagram. We excluded all obviously transient LMXBs and IMXBs since, in most
cases, it is unclear how to estimate the long-term average X-ray
luminosity\footnote{We note that the issue of determining the
long-term luminosity of an X-ray source is quite difficult -- even for
so called ``steady sources''. The entire history of X-ray astronomy is
shorter than 40 years; while the typical timesteps in our binary
evolution code may range from $10^2 - 10^7$ years. Therefore, even
X-ray sources which appear steady over the entire history of X-ray
astronomy may, in fact, be transient over the longer term, e.g.,
comparable to the time steps in our evolution code.}. We then selected
16 LMXBs and IMXBs (i) whose orbital periods are known, (ii) whose
X-ray luminosities do not vary wildly, and (iii) where a distance
to the source could be estimated. These are shown overplotted on
Figure 18; they include 2 ``Z sources'' (triangles), 8 ``atoll
sources''(squares), 3 X-ray pulsars (stars), and 3 ``accretion disk
corona sources'' (circles). (For references see, e.g.,
\markcite{Paradijs2}van Paradijs 1995; \markcite{Swank}Christian
\& Swank 1997.)
For the latter group of sources, the observed X-ray flux is thought to
be severely affected by an accretion disk corona, and therefore the
inferred value of $\dot M$ is shown only as a lower limit. We used a
simple factor of $10^{-8} M_{\odot}\, {\rm yr}^{-1} \equiv 10^{38}$
erg\,s$^{-1}$ in converting X-ray luminosity to mass-accretion rate.

The first obvious fact in comparing the theoretical probability
distribution to the locations of known LMXBs and IMXBs in the $\dot
M-P_{\rm orb}$ plane (Fig.~18) is that only a relative handful lie
plausibly in or near the shaded region. In fact, 10 of the 16 sources
lie at or outside the $1\%$ upper and lower probability contours. The
largest discrepancies come from luminous LMXBs with shorter orbital
periods (i.e., $\la$ 1 day). There are several important caveats to
note before viewing the comparison made in Figure 18 as being grossly
discrepant. First, as mentioned above, the evolution tracks that went
into the production of the probability contours in Figure 18 are not
weighted by the relative probabilities of achieving their initial
binary configurations in nature. Second, no transient X-ray sources
have been included in the figure. Many of these sources probably have
mean values of $\dot{M}$ of $\la 3 \times 10^{-10} M_{\odot}\, {\rm
yr}^{-1}$ (and possibly even higher for the larger values of $P_{\rm
orb}$; \markcite{Paradijs}van Paradijs 1996; \markcite{King4}King,
Kolb, \& Sienkiewicz 1997) which cover a substantial portion of the
shaded (high probability) region. Third, there are serious
observational selection effects to consider, in that it is generally
true that the most luminous X-ray sources are studied in detail,
yielding higher probabilities of optical identifications which, in
turn, can lead to orbital period determinations. At least the first
two of these shortcomings of Figure 18 will be addressed in our binary
population synthesis study (\markcite{Pfahl}Pfahl et al.\ 2001).

One potentially very important effect that has not been included in
our binary calculations is the effect of X-ray irradiation on the
secondary which could significantly alter the evolution of these
systems and increase the mass-accretion rate by either driving a
strong wind from the secondary (\markcite{Ruderman2}Ruderman et al.\
1989) or by causing significant expansion of the secondary
(\markcite{Podsi3}Podsiadlowski 1991;
\markcite{Harpaz}Harpaz \& Rappaport 1991). While these irradiation
effects are still poorly understood, even a relatively moderate
irradiation-driven expansion of the secondary may cause mass-transfer
cycles (Hameury et al.\ 1993) characterized by relatively short phases
of enhanced mass transfer and long detached phases. During the X-ray
phases these systems would appear to be much more luminous than
without the inclusion of X-ray irradiation effects. We also plan to
examine this possibility in our BPS study.

\section{Application to Binary Millisecond Pulsars}

There are currently about 1400 radio pulsars known
(\markcite{Taylor}Taylor, Manchester, \& Lyne 1993; V. Kaspi 2001,
private communication).  Of these, $\sim 100$ have at least one of the
following properties (V. Kaspi 2001, private communication): (i) a very
short pulse period ($\sim 77$ with $P \lesssim 12\,$ms); (ii) a
relatively weak magnetic field ($\sim 46$ with $B \lesssim 10^{10}\,$G);
(iii) membership in a binary system ($\sim 66$); and/or (iv) location
in a globular cluster ($\sim 45$).  These systems are widely believed
to be ``recycled" pulsars, i.e., NSs whose magnetic field has decayed
away and which have been spun up to high rotation rates by the
accretion of matter from a companion star (see, e.g.,
\markcite{Bhatta}Bhattacharya \& van den Heuvel 1991).  In the
Galactic plane, there are several distinct classes of binary radio
pulsars.  One major class involves systems with low-mass companions
($0.10\,$--$\, 0.4\, M_\odot$) and nearly circular orbits.  These range in
$P_{\rm orb}$ from a fraction of a day to 1000 days.  There is a
dearth of these pulsars in the period range of 12 and 68 days.  The
masses of most of the companions to these pulsars are known only
approximately from the measured mass functions.  Based on the scenario
for their formation, which involves stable mass transfer from a
low-mass giant, there is a theoretically predicted relation between
the orbital period of these systems and the mass of the remnant
companion white dwarf (see, e.g., \markcite{Rappaport1}Rappaport et
al.\ 1995).  In fact, the locus of points in Figure 12 tracing the
maximum value of $P_{\rm orb}$ at any given white dwarf mass matches
the theoretical relation rather closely (see also \S~4.1).  Most of
the model systems helping to define this relation, however, have
orbital periods between $\sim$12 and 120 days -- at least the first
half of which fall in the period ``gap'' found observationally.  There
is also another cluster of model systems with $P_{\rm orb}$ between
$\sim$11 and 85 minutes; these are of shorter periods than any of the
binary pulsars discovered thus far in the Galactic disk.  Again we
note the caveat discussed in \S~5 that our library of binary
models has not been weighted according to the probability of achieving
their initial binary parameters at the onset of mass transfer, e.g.,
in the context of a full binary population synthesis calculation.

Another class of binary radio pulsars are the ones with substantially
more massive white dwarf companions which distinctly do not fit the
scenario described above (with a low-mass giant donor) and do not lie
in the $P_{\rm orb}\,$--$\,M_{\rm wd}$ plane near the associated theoretical
relationship.  These systems also have nearly circular orbits and
$P_{\rm orb}$ in the range of $\sim 1\,$--\,10 days. It has been proposed
for some time now that these systems result from donor stars which are
more massive than the neutron star, thereby leading to unstable mass
transfer and a common envelope phase (see, e.g., \markcite{Taam}Taam
\& van den Heuvel 1986).  Our models with donors that are initially of
intermediate mass naturally lead to this type of system {\it without}
a common envelope phase (see also \S~4 and \markcite{Tauris2}Tauris
et al.\ 2000).  Such model systems are found in abundance in Figure~13
(triangles situated well below and to the right of the theoretical curve).

Yet another class of binary radio pulsars are systems that contain
planetary mass companions (i.e., $M \la 0.02 M_\odot$) which are in
the process of being ablated by the radiation from the pulsar (e.g.,
1957+20; \markcite{Fruchter}Fruchter, Stinebring, \& Taylor 1988).  In
this regard, we note that in our binary evolution calculations, the
mass transfer is allowed to continue until either the donor star
becomes detached from its Roche lobe or its non-degenerate envelope
has been completely stripped.  Of course, as the neutron star accretes
matter from the companion it will be spun up by accretion torques --
the maximum spin period being determined by a combination of, $\dot
M$, the total mass accreted, and the strength of the neutron star's
magnetic field.  For weak surface magnetic fields (i.e., $\la
10^9\,$G), the accretion-induced spin period is given approximately by
$P = 3.5\times (\Delta M/0.01 M_\odot)^{-1/2}$~ms, where $\Delta M$ is the
accreted mass.  For higher magnetic fields, the minimum accretion
induced rotation period scales as $1.9 B_{9}^{6/7}$ ms, for an
Eddington-limited luminosity, where $B_9$ is the surface dipole field
strength in units of $10^9\,$G (see, e.g.,
\markcite{Bhatta}Bhattacharya \& van den Heuvel 1991).  Thus, at some
point in the evolution, prior to the exhaustion of the donor's
envelope, a spun-up neutron star may turn on as a radio pulsar.  This
could have two important consequences for the subsequent evolution of
the binary.  First, the pulsar radiation (in the form of both
electromagnetic waves and a relativistic wind) may exert sufficient
pressure on the incoming accretion flow that all further accretion is
halted.  Second, the pulsar radiation may ablate material from the
donor star, and in some cases possibly evaporate it altogether (see,
e.g., \markcite{Ruderman1}Ruderman, Shaham, \& Tavani 1989;
\markcite{Heuvel}van den Heuvel \& van Paradijs 1988,
\markcite{Bhatta}Bhattacharya \& van den Heuvel 1991).  In the
present binary evolution calculations, we do not take either a
possible pulsar turnon into account or the subsequent effects of the
pulsar radiation on the donor star.  In our BPS study we will attempt to
use simplified prescriptions to handle both of these processes, i.e.,
pulsar turnon and ablation of the donor, although there are obviously
still many uncertainties concerning  both of these.

The final class of binary pulsars we comment on consists of a
pair of neutron stars.  While these systems are important for
exploring binary evolution, acting as laboratories for general relativity,
and yielding potentially detectable gravity wave
signals when they merge, our study does not shed any new light
on their formation.  This results from the fact that our
highest mass donor stars are $7~M_\odot$ which is too low to
form a second neutron star.

Many of the same classes of binary radio pulsars that are found in the
plane have also been discovered in globular clusters.  In particular,
there are 22 radio pulsars known in 47 Tuc (\markcite{Camilo}Camilo et
al.\ 2000), 8 in M15 (\markcite{Anderson}Anderson 1992), and 2 each in
M5, M13, Ter 5, and NGC 6624.  At least 10 of the radio pulsars in 47
Tuc are in binary systems with periods ranging from 1.5 hr to 2 days.
This abundance of recycled pulsars in globular clusters is widely
attributed to the dense stellar environment which can
lead to 2-, 3-, and 4-body stellar encounters at interestingly high rates.
Thus, through a combination of processes such as 2-body tidal capture (e.g.,
\markcite{Fabian}Fabian, Pringle, \& Rees 1975; \markcite{Distefano2}
Di Stefano \& Rappaport 1992) and exchange interactions where a field
neutron star replaces a normal star in a binary, numerous neutron star
binaries should be formed (see, e.g., \markcite{Rasio}Rasio, Pfahl, \&
Rappaport 2000; \markcite{Rappaport5}Rappaport et al. 2001).

We have already discussed in \S~4.2 how the globular-cluster X-ray
sources fit our evolution scenarios.  However, we need to
check if the binary millisecond pulsars also fit naturally into the
same evolutionary scenarios.  In 47 Tuc, there is a group (``A'') of
five pulsars which have masses of $0.02-0.03~M_\odot$ and periods of
1.5\,--\,5.5\,hr, while a second group (``B'') has masses approximately 10
times
higher and periods of 0.1\,--\,2\,d.   The evolutionary scenarios for
producing
the ultracompact X-ray binaries discussed in \S 4.2 may also provide a possible
path to the formation of binary radio pulsars of the type found in the
47 Tuc group A pulsars.  In the process of evolving to very short periods
(11--83 min)., these mass transfer binaries will naturally pass through
orbital periods of 1.5--5 hr.  However, at these periods the donor masses
are substantially larger than the typical values of $\sim 0.025 M_\odot$
found for the companions in the group A systems.  On the other hand, if
at some point in the binary evolution, the neutron star has been spun up to
msec periods, pulsar radiation may turn on and both shut off further mass
transfer and ablate the donor star until it has been reduced to planetary mass.
It is also true that after the minimum period is reached in the ultracompact
systems (see \S 4.2), the orbit will, in principle, expand back into the range
of about an hour or so within a Hubble time.  In order for such systems to
return all the way back to periods of 1.5--5 hours, some other effect, such as
sustained tidal heating, would be required (as is also invoked for the case
of the post-common envelope scenario proposed by \markcite{Rasio}Rasio
et al.\ 2000). A characteristic of our ``ultracompact'' evolutionary
scenario is that the correct initial orbital periods between $\sim13-18$ hr
arise naturally from tidal capture in globular clusters.  By contrast, the
common envelope scenario proposed by \markcite{Rasio}Rasio et al.\ (2000)
follows more naturally from the wider orbits left by 3-body encounters.

The orbital period range of $\sim 0.1-2$ days for the group B pulsars
is traversed in many of our binary evolutions (see, e.g., Fig.~2a).  However,
in all but 2 cases, the masses of the donor stars in this period range are
much higher than the group B pulsar companions.  In addition, the
donor stars at this phase of the evolution are still quite H-rich and
would therefore not resemble the inferred He white dwarf companions of
these pulsars.  Again, it is possible that these evolutions are interrupted by
the turn-on of a strong radio pulsar at just the right values of $P_{\rm orb}$
to match the group B systems.  At present, we do not have a good explanation
for how this would happen, nor do we know the reason for the existence
of two rather distinct groups of msec pulsars.

A long-standing problem in our understanding of millisecond pulsars,
known as the birthrate problem, is that, in the standard model of
LMXBs, the birthrate of LMXBs appears to be a factor of 10 to 100
lower than the birthrate of millisecond pulsars, which are believed to
be their direct descendants. This problem exists for millisecond
pulsars both in the Galactic disk (\markcite{Kulkarni}Kulkarni
\& Narayan 1988; \markcite{Johnston}Johnston \& Bailes 1991) and
in globular clusters (\markcite{Fruchter2} Fruchter \& Goss 1990;
\markcite{Kulkarni2} Kulkarni, Narayan, \& Romani 1990).
This discrepancy may be the result of an overestimate of the
LMXB lifetime, typically taken to be $\sim 5\times 10^9\,$yr in these
estimates. If a large fraction of X-ray binaries are IMXBs, one might
expect that the time these systems spend as X-ray emitters could be
significantly reduced, which would then alleviate the
problem. However, our calculations show that even IMXBs spend most
most of their X-ray active lifetime as low-mass systems and that
consequently the duration of the X-ray active lifetime is generally
not much lower than for true LMXBs (see the column $\Delta
t_{\dot{M}}$ in Table~A1). Thus the inclusion of IMXBs does not
immediately solve the birthrate problem. This problem may
also be related to the problem of the low median X-ray luminosities
found in our calculations (as discussed in \S~5) and may have a
similar resolution:  if irradiation-driven mass-transfer cycles
operate for low-mass systems (see \S~5), these would not only increase
the mass-transfer rates during the X-ray active portion of the cycles but
also reduce the duration of the X-ray active lifetime of these
systems by a proportionate amount. This could provide a simultaneous
solution to both of these problems, the X-ray luminosity and the
birthrate problem.

As we have shown, our binary evolution models pertain directly to the
binary radio pulsars found in both the Galactic plane and in globular
clusters.  However, before we can draw definitive conclusions about the
relative and absolute populations of the different classes of
these objects, in both the plane and in clusters, we must await the results
of our binary population synthesis study (\markcite{Pfahl}Pfahl et al.\
2001).

\section{Summary and Outlook}

As this study has shown, the evolution of low- and intermediate-mass
X-ray binaries is much more complex than previously believed, and the
standard model for these systems where mass transfer is driven only by
magnetic braking, gravitational radiation, and occasionally nuclear
evolution is the exception rather than the rule. Our evolutionary
sequences show an enormous variety of evolutionary channels which may
explain the large diversity in observed systems.  Indeed many of the
best-studied systems in the Galactic disk (e.g., Her X-1, Cyg X-2, Sco
X-1, GRO J1744-28, 4U 1626-67 and 4U 1915-05) as well as all globular
clusters sources with known orbital periods can be identified with
particular sequences in our library of models.  This demonstrates
the importance of these results for our understanding of X-ray
binaries and millisecond pulsars, believed to be their descendants.

However, our results also show that there are a number of problems still
remaining, in particular the low median X-ray luminosities as compared to the
luminosity of well observed systems, the orbital-period distribution
of millisecond pulsars, and the millisecond pulsar birthrate problem. To shed
more light on the significance of these discrepancies, we have
initiated a systemic binary population synthesis study
(\markcite{Pfahl}Pfahl et al.\ 2001) where we implement this library
of models in a population synthesis code. This will not only allow us
to quantify these discrepancies more precisely, but also to examine
possible solutions (e.g., irradiation-driven cycles, pulsar turnon,
pulsar evaporation, etc.)  and should ultimately help us to improve
our understanding of the evolution X-ray binaries and the formation of
millisecond pulsars.

\section*{Acknowledgments}

We would like to thank Ron Remillard, Mike Muno, and Phil Charles for
providing important observational information on several LMXBs and many
useful discussions.
This work was in part supported by the National Aeronautics and Space
Administration under ATP grant NAG5-8368.

\appendix

In this appendix, we present in tabular form (Table~A1) some of the main
characteristics of the 100 binary sequences in this study, where the  selected
parameters depend on the type of evolution encountered. In all sequences,
the initial mass of the compact object, assumed to be a neutron star,
was taken to be 1.4\Ms. The initial mass of the secondary, $M_2$, ranges
from 0.6 to 7\Ms. For these masses, the table first lists (in the first row)
the initial parameters for each sequence: the initial central hydrogen
mass fraction, $X_c^i$; the initial fractional mass of the H-exhausted
core, $M_c^i/M$; the age of the secondary at the beginning of mass transfer,
$t^i\,$(yr). The next parameters give the key binary parameters at
the end of each calculation and the type of end product, where $M_2^f$
and $M_1^f$ are the final masses of the secondary and primary (in \Ms),
respectively, $P^f$ is the final orbital period (in d), $\Delta t$ gives
the total time since the beginning of mass transfer (in yr) and `Type'
indicates the type of end product (`short': compact system; `He': wide
system with a He white dwarf secondary; `HeCO': wide system with a
HeCO white dwarf secondary; `del dyn': delayed dynamical
instability; `dyn': dynamical mass transfer; `?' indicates that the system
may be dynamically unstable).

Note that the calculations were terminated at different points for the
different types of evolution. For the ultracompact systems, the
calculations were generally terminated just after the period minimum
when the secondaries have become fully degenerate. In systems where
the secondary becomes a He or a HeCO white dwarf, the calculations are
continued either up to the point where the secondary has settled on
the cooling sequence for degenerate stars or until the beginning of the
first hydrogen shell flash (although in many cases, we continued the
calculations through all flashes). In systems that experience a
dynamical instability, the calculations were generally terminated when the
secondaries overfilled their Roche lobes by a factor of 1.5. In cases,
where the maximum overflow factor was less than 1.5, we continued the
calculations (pretending that the systems did not experience a
spiral-in phase) just as for the systems that were dynamically stable
(these are the systems marked with `?' in the `Type' column).

The next three parameters in the first row give the average and the
maximum mass-transfer rate, $<\dot{M}>$ and $\dot{M}_{\rm max}$ (in \Msyr),
respectively, and the total duration of the mass-transfer phase for
each sequence, $\Delta t_{\dot{M}}$ (in yr), where we considered only
phases where $\dot{M}$ exceeded a rate of $10^{-12}\Msyr$. The last column
in the first row indicates whether the secondary experienced hydrogen
shell flashes before settling onto the sequence for degenerate stars.

In the second row, the first four parameters apply to systems that
evolve towards short periods: column `gap' gives the range of orbital
periods (in hr) for the sequences that experienced a period gap;
$M_2^{\rm min}$ gives the secondary mass (in \Ms) at the period minimum,
$P_{\rm min}$ (in min), and $X_s^{\rm min}$ the surface hydrogen
abundance at the point.

The last 6 columns apply to systems that (may) experience dynamical
mass transfer, where $f_{\rm over}$ is the maximum overflow factor
(i.e., the ratio of the secondary radius to the Roche-lobe radius),
$t_{\rm atm}$ and $t_{\rm rad}$ give the duration
of phases with `atmospheric' and `radiative' Roche-lobe overflow
(in kyr), respectively, before the onset of the dynamical instability.
Finally, $M_2^{\rm dyn}$ and $M_1^{\rm dyn}$ are the secondary and
primary masses (in \Ms), respectively, and $P^{\rm dyn}$ is the orbital
period (in d) when the mass-transfer rate starts to exceed
$10^{-4}\Msyr$, which is close to the point where many/most of these systems
are expected to become dynamically unstable.


\newpage

%
%


\begin{deluxetable}{cccccccccccccc}
\tabletypesize{\scriptsize}
\tablecolumns{14}
\tablewidth{0pt}
\tablenum{A1}
\tablecaption{Selected Properties of Binary Sequences}
\tabcolsep=2pt
\tablehead{&$P^i\,$(d)&$X_c^i$&$M_c^i/M$&$t^i\,$(yr)&$M_2^f\,$($M_\odot$)&
            $M_1^f\,$($M_\odot$)&$P^f\,$(d)&$\Delta t\,$(yr)&Type&
            $<\dot{M}>$\tablenotemark{a}&
	     $\dot{M}_{\rm max}$\tablenotemark{a}&
	     $\Delta t_{\dot{M}}\,$(yr)&H flash?\\
\tablevspace{2pt}
&&&&gap (hr)&$M_2^{\rm min}\,$($M_\odot$)&$P_{\rm min}\,$(min)&
$X_s^{\rm min}$&$f_{\rm over}$&$t_{\rm atm}\,$(kyr)&$t_{\rm rad}$\,(kyr)&
$M_2^{\rm dyn}$\,($M_\odot$)&$M_1^{\rm dyn}$\,($M_\odot$)&$P^{\rm dyn}\,$(d)
\\
\tablevspace{-3pt}
}
\startdata
\\
\tablevspace{-10pt}
\multicolumn{13}{l}{$M_2^i =  0.6\Ms$:}\\
\tablevspace{6pt}
&  0.18& 0.685& 0.000&0.00E+00&
 0.048& 1.676&   0.060&4.87E+09&
short  & 1.2E-10& 4.1E-09& 4.8E+09& n\\
&&&&2.86\,--\,3.04& 0.066& 82.4& 0.68\\
\tablevspace{0pt}
\multicolumn{14}{l}{$M_2^i =  0.8\Ms$:}\\
\tablevspace{6pt}
&  0.24& 0.685& 0.000&3.00E+06&
 0.047& 1.776&   0.059&4.86E+09&
short  & 1.6E-10& 5.2E-09& 4.6E+09& n\\
&&&&2.76\,--\,3.03& 0.067& 80.3& 0.68\\
&  0.26& 0.436& 0.000&7.99E+09&
 0.048& 1.776&   0.057&3.62E+09&
short  & 2.1E-10& 4.2E-09& 3.6E+09& n\\
&&&&2.07\,--\,2.23& 0.077& 77.6& 0.64\\
\tablevspace{0pt}
\multicolumn{14}{l}{$M_2^i =  1.0\Ms$:}\\
\tablevspace{6pt}
&  0.30& 0.685& 0.000&0.00E+00&
 0.049& 1.875&   0.060&4.65E+09&
short  & 2.1E-10& 6.0E-09& 4.6E+09& n\\
&&&&2.80\,--\,3.07& 0.066& 83.2& 0.68\\
&  0.35& 0.323& 0.000&4.69E+09&
 0.048& 1.876&   0.058&3.75E+09&
short  & 2.6E-10& 5.1E-09& 3.7E+09& n\\
&&&&2.26\,--\,2.41& 0.062& 79.8& 0.62\\
&  0.52& 0.000& 0.000&9.53E+09&
 0.034& 1.883&   0.041&3.01E+09&
short  & 3.2E-10& 4.3E-09& 3.0E+09& n\\
&&&&     none     & 0.055& 48.2& 0.34\\
&  0.74& 0.000& 0.055&1.11E+10&
 0.194& 1.803&   1.662&1.60E+10&
He     & 3.8E-10& 3.9E-09& 2.1E+09& y\\
&&&&&\\
&  1.07& 0.000& 0.125&1.16E+10&
 0.246& 1.777&   8.863&9.31E+08&
He     & 9.4E-10& 6.2E-09& 8.1E+08& y\\
&&&&&\\
&  4.59& 0.000& 0.202&1.20E+10&
 0.307& 1.744&  52.311&1.44E+08&
He     & 5.4E-09& 2.7E-08& 1.3E+08& y\\
&&&&&\\
& 24.69& 0.000& 0.277&1.22E+10&
 0.360& 1.660& 208.733&3.27E+07&
He     & 4.1E-08& 9.7E-07& 1.6E+07& y\\
&&&&&\\
&102.70& 0.000& 0.341&1.23E+10&
 0.430& 1.476& 617.434&5.00E+08&
He     & 1.3E-07& 2.2E-06& 4.5E+06& n\\
&&&&&\\
\tablebreak
\tablevspace{0pt}
\multicolumn{14}{l}{$M_2^i =  1.2\Ms$:}\\
\tablevspace{6pt}
&  0.39& 0.671& 0.000&1.79E+08&
 0.049& 1.976&   0.060&4.66E+09&
short  & 2.5E-10& 5.8E-09& 4.6E+09& n\\
&&&&2.54\,--\,2.77& 0.066& 83.0& 0.67\\
&  0.54& 0.385& 0.000&3.45E+09&
 0.040& 1.980&   0.056&3.40E+09&
short  & 3.4E-10& 4.4E-09& 3.4E+09& n\\
&&&&     none     & 0.055& 77.8& 0.53\\
&  0.72& 0.099& 0.000&5.34E+09&
 0.025& 1.987&   0.035&3.04E+09&
short  & 4.0E-10& 6.2E-09& 3.0E+09& n\\
&&&&     none     & 0.071& 39.0& 0.12\\
&  1.00& 0.000& 0.049&5.77E+09&
 0.248& 1.876&  10.037&8.14E+08&
He     & 1.3E-09& 6.5E-09& 7.4E+08& y\\
&&&&&\\
&  1.42& 0.000& 0.128&5.91E+09&
 0.268& 1.866&  19.391&5.27E+08&
He     & 2.0E-09& 1.8E-08& 4.6E+08& y\\
&&&&&\\
&  2.01& 0.000& 0.147&6.34E+09&
 0.282& 1.859&  28.880&3.76E+08&
He     & 2.8E-09& 3.8E-08& 3.4E+08& y\\
&&&&&\\
&  3.41& 0.000& 0.164&7.07E+09&
 0.304& 1.840&  49.271&2.02E+08&
He     & 5.3E-09& 3.7E-08& 1.7E+08& y\\
&&&&&\\
&  5.82& 0.000& 0.182&7.84E+09&
 0.323& 1.817&  79.175&1.09E+08&
He     & 1.0E-08& 3.6E-08& 8.4E+07& y\\
&&&&&\\
\tablevspace{0pt}
\multicolumn{14}{l}{$M_2^i =  1.4\Ms$:}\\
\tablevspace{6pt}
&  0.47& 0.674& 0.000&7.17E+07&
 0.075& 2.062&   0.009&1.01E+10&
short  & 1.3E-10& 2.2E-08& 1.0E+10& n\\
&&&&     none     & 0.106& 10.8& 0.01\\
&  0.70& 0.347& 0.000&2.20E+09&
 0.050& 2.075&   0.013&4.65E+09&
short  & 2.9E-10& 1.5E-08& 4.6E+09& n\\
&&&&     none     & 0.099& 12.0& 0.02\\
&  0.96& 0.102& 0.000&3.16E+09&
 0.211& 1.994&   2.677&0.00E+00&
He     & 1.1E-09& 6.6E-09& 1.1E+09& y\\
&&&&&\\
&  1.91& 0.000& 0.114&3.49E+09&
 0.296& 1.895&  37.896&2.04E+08&
He     & 9.6E-09& 9.9E-08& 1.1E+08& y\\
&&&&&\\
&  3.76& 0.000& 0.138&3.52E+09&
 0.315& 1.857&  64.060&1.52E+08&
He     & 1.1E-08& 2.9E-07& 9.9E+07& y\\
&&&&&\\
&  6.42& 0.000& 0.169&3.57E+09&
 0.335& 1.822&  98.308&7.18E+07&
He     & 2.0E-08& 4.6E-07& 5.5E+07& y\\
&&&&&\\
\tablebreak
\tablevspace{0pt}
\multicolumn{14}{l}{$M_2^i =  1.6\Ms$:}\\
\tablevspace{6pt}
&  0.49& 0.673& 0.000&7.02E+07&
 0.027& 2.186&   0.035&6.29E+09&
short  & 2.6E-10& 3.2E-09& 6.3E+09& n\\
&&&&     none     & 0.061& 35.2& 0.31\\
&  0.83& 0.355& 0.000&1.48E+09&
 0.264& 2.068&  17.644&1.59E+09&
He     & 9.5E-10& 4.1E-09& 1.4E+09& y\\
&&&&&\\
&  1.30& 0.100& 0.000&2.13E+09&
 0.287& 2.056&  31.520&3.17E+08&
He     & 5.1E-09& 1.8E-08& 2.6E+08& y\\
&&&&&\\
&  1.86& 0.000& 0.046&2.30E+09&
 0.306& 1.885&  40.234&9.45E+07&
He     & 2.4E-08& 9.9E-08& 5.3E+07& y\\
&&&&&\\
&  2.73& 0.000& 0.112&2.32E+09&
 0.318& 1.765&  55.029&7.05E+07&
He     & 3.4E-08& 5.0E-07& 3.8E+07& y\\
&&&&&\\
&  4.63& 0.000& 0.128&2.32E+09&
 0.335& 1.738&  82.336&5.87E+07&
He     & 4.0E-08& 2.0E-06& 3.1E+07& y\\
&&&&&\\
&  9.07& 0.000& 0.146&2.34E+09&
 0.359& 1.696& 136.644&1.24E+09&
He     & 6.1E-08& 6.9E-06& 2.2E+07& y\\
&&&&&\\
& 16.62& 0.000& 0.162&2.34E+09&
 0.383& 1.636& 218.570&2.94E+07&
He     & 7.8E-08& 1.9E-05& 1.6E+07& y\\
&&&&&\\
\tablevspace{0pt}
\multicolumn{14}{l}{$M_2^i =  1.8\Ms$:}\\
\tablevspace{6pt}
&  0.50& 0.662& 0.000&9.58E+07&
 0.050& 2.275&   0.019&6.35E+09&
short  & 2.8E-10& 2.5E-08& 6.4E+09& n\\
&&&&     none     & 0.061& 25.3& 0.13\\
&  0.97& 0.330& 0.000&1.11E+09&
 0.277& 2.152&  25.739&1.15E+09&
He     & 1.5E-09& 5.2E-08& 1.0E+09& y\\
&&&&&\\
&  1.66& 0.099& 0.000&1.49E+09&
 0.298& 2.099&  44.217&1.90E+08&
He     & 9.0E-09& 5.7E-08& 1.8E+08& y\\
&&&&&\\
&  3.29& 0.000& 0.104&1.62E+09&
 0.320& 1.660&  71.218&3.05E+07&
He     & 9.2E-08& 7.9E-05& 1.7E+07& y\\
&&&&&\\
&  6.60& 0.000& 0.127&1.62E+09&
 0.359& 1.614& 109.041&2.80E+07&
He     & 8.1E-08& 8.7E-06& 1.6E+07& y\\
&&&&&\\
\tablebreak
\tablevspace{0pt}
\multicolumn{14}{l}{$M_2^i =  2.1\Ms$:}\\
\tablevspace{6pt}
&  0.50& 0.683& 0.000&2.00E+06&
 0.037& 2.341&   0.045&6.14E+09&
short  & 3.4E-10& 1.1E-07& 6.0E+09& n\\
&&&&     none     & 0.049& 62.7& 0.32\\
&  0.71& 0.489& 0.000&4.40E+08&
 0.231& 2.198&   6.333&2.65E+09&
He     & 7.6E-10& 1.4E-07& 2.5E+09& y\\
&&&&&\\
&  1.10& 0.300& 0.000&7.26E+08&
 0.254& 2.135&  19.660&9.35E+08&
He     & 2.2E-09& 2.3E-07& 8.8E+08& y\\
&&&&&\\
&  2.01& 0.096& 0.000&9.29E+08&
 0.322& 2.167&  39.047&1.63E+08&
HeCO   & 1.1E-08& 1.7E-07& 1.6E+08& y\\
&&&&&\\
&  2.87& 0.000& 0.039&9.99E+08&
 0.360& 1.614&  50.885&1.41E+08&
HeCO   & 1.4E-07& 5.8E-07& 1.3E+07& y\\
&&&&&\\
&  4.42& 0.000& 0.101&1.00E+09&
 0.367& 1.529&  71.877&1.33E+08&
HeCO   & 1.8E-07& 1.0E-04& 7.8E+06& y\\
&&&&&\\
& 28.95& 0.000& 0.150&1.01E+09&
 0.562& 1.434& 173.217&1.66E+08&
HeCO?  & 4.8E-07& 1.0E-04& 1.8E+06& y\\
&&&&&&&&1.09&&  85.2& 1.872& 1.4005& 28.68\\
\tablevspace{0pt}
\multicolumn{14}{l}{$M_2^i =  2.4\Ms$:}\\
\tablevspace{6pt}
&  0.51& 0.676& 0.000&1.74E+07&
 0.031& 2.270&   0.027&6.35E+09&
short  & 3.7E-10& 3.7E-07& 6.3E+09& n\\
&&&&     none     & 0.058& 29.0& 0.18\\
&  1.06& 0.327& 0.000&4.80E+08&
 0.225& 2.077&   8.464&1.27E+09&
He     & 1.8E-09& 6.7E-07& 1.2E+09& y\\
&&&&&\\
&  2.08& 0.097& 0.000&6.39E+08&
 0.340& 2.021&  21.018&1.98E+08&
HeCO   & 1.1E-08& 7.9E-07& 1.9E+08& y\\
&&&&&\\
&  4.22& 0.000& 0.065&6.86E+08&
 0.421& 1.495&  47.379&1.29E+08&
HeCO   & 3.7E-07& 2.5E-06& 5.3E+06& y\\
&&&&&\\
&  8.56& 0.000& 0.116&6.88E+08&
 0.429& 1.460&  89.865&1.20E+08&
HeCO?  & 3.4E-07& 1.0E-04& 3.4E+06& y\\
&&&&&&&&1.12&&  24.8& 2.352& 1.4002&  8.43\\
\tablebreak
\tablevspace{0pt}
\multicolumn{14}{l}{$M_2^i =  2.7\Ms$:}\\
\tablevspace{6pt}
&  0.52& 0.679& 0.000&8.43E+06&
 0.031& 2.197&   0.045&6.40E+09&
short  & 4.2E-10& 9.7E-07& 6.4E+09& n\\
&&&&     none     & 0.059& 55.7& 0.41\\
&  1.09& 0.328& 0.000&3.46E+08&
 0.231& 1.995&  10.106&1.42E+09&
He     & 1.8E-09& 1.6E-06& 1.4E+09& y\\
&&&&&\\
&  2.14& 0.100& 0.000&4.59E+08&
 0.358& 1.856&  18.993&6.88E+08&
HeCO   & 1.4E-08& 2.1E-06& 1.6E+08& n\\
&&&&&\\
&  4.27& 0.000& 0.067&4.93E+08&
 0.483& 1.454&  29.826&1.08E+08&
HeCO   & 7.1E-07& 4.6E-06& 3.1E+06& y\\
&&&&&\\
&  8.52& 0.000& 0.112&4.94E+08&
 0.490& 1.437&  57.094&1.06E+08&
HeCO?  & 5.5E-07& 1.0E-04& 2.1E+06& y\\
&&&&&&&&1.13&&  10.3& 2.506& 1.4001&  7.76\\
\tablevspace{0pt}
\multicolumn{14}{l}{$M_2^i =  3.0\Ms$:}\\
\tablevspace{6pt}
&  0.54& 0.669& 0.000&1.66E+07&
 0.025& 2.140&   0.036&9.45E+09&
short  & 3.2E-10& 2.0E-06& 9.4E+09& n\\
&&&&     none     & 0.065& 33.9& 0.19\\
&  1.10& 0.340& 0.000&2.51E+08&
 0.259& 1.909&  23.583&1.05E+09&
He     & 2.7E-09& 3.3E-06& 1.0E+09& y\\
&&&&&\\
&  2.23& 0.098& 0.000&3.41E+08&
 0.387& 1.716&  15.191&5.75E+08&
HeCO   & 2.1E-08& 4.3E-06& 1.2E+08& n\\
&&&&&\\
&  4.46& 0.000& 0.068&3.66E+08&
 0.546& 1.440&  19.257&3.87E+08&
HeCO   & 1.2E-06& 7.4E-06& 2.1E+06& n\\
&&&&&\\
&  8.95& 0.000& 0.091&3.66E+08&
 0.555& 1.433&  37.376&9.65E+08&
HeCO   & 1.4E-06& 2.0E-05& 1.8E+06& n\\
&&&&&\\
& 17.52& 0.000& 0.122&3.66E+08&
 0.561& 1.421&  71.054&2.70E+08&
HeCO?  & 5.5E-07& 1.0E-04& 1.1E+06& n\\
&&&&&&&&1.41&&  13.6& 2.963& 1.4001& 17.12\\
\tablebreak
\tablevspace{0pt}
\multicolumn{14}{l}{$M_2^i =  3.5\Ms$:}\\
\tablevspace{6pt}
&  0.55& 0.684& 0.000&6.20E+05&
 0.049& 1.903&   0.060&4.46E+09&
short? & 4.9E-10& 1.0E-04& 4.3E+09& n\\
&&&&&&&&1.11&2030.8& 334.7& 2.350& 1.4077&  0.26\\
&  1.14& 0.337& 0.000&1.62E+08&
 0.235& 1.831&  11.030&1.36E+09&
He     & 2.5E-09& 8.6E-06& 1.3E+09& y\\
&&&&&\\
&  2.28& 0.103& 0.000&2.17E+08&
 0.433& 1.624&   9.070&2.45E+09&
HeCO   & 4.0E-08& 1.2E-05& 7.5E+07& n\\
&&&&&\\
&  4.78& 0.000& 0.078&2.33E+08&
 0.658& 1.423&   9.133&4.20E+08&
HeCO   & 3.6E-06& 2.1E-05& 7.7E+05& n\\
&&&&&\\
& 11.03& 0.000& 0.101&2.33E+08&
 0.672& 1.418&  20.118&3.69E+09&
HeCO   & 3.1E-06& 5.8E-05& 9.2E+05& n\\
&&&&&\\
& 22.37& 0.000& 0.125&2.34E+08&
 2.719& 1.400&  12.665&1.01E+06&
Dyn    & 4.5E-06& 1.0E-04& 4.8E+03& n\\
&&&&&&&&1.50&&   4.8& 3.468& 1.4000& 21.91\\
\tablevspace{0pt}
\multicolumn{14}{l}{$M_2^i =  4.0\Ms$:}\\
\tablevspace{6pt}
&  0.57& 0.676& 0.000&4.47E+06&
 2.118& 1.404&   0.161&2.04E+06&
Del Dyn& 7.3E-07& 1.0E-04& 9.6E+05& n\\
&&&&&&&&1.50& 848.1& 114.8& 3.220& 1.4033&  0.28\\
&  1.17& 0.338& 0.000&1.19E+08&
 0.237& 1.757&  12.122&1.35E+09&
He?    & 5.1E-09& 1.0E-04& 4.0E+08& y\\
&&&&&&&&1.20& 410.3&  84.2& 2.921& 1.4026&  0.47\\
&  2.32& 0.102& 0.000&1.59E+08&
 0.466& 1.573&   5.776&6.29E+07&
HeCO   & 1.3E-07& 1.7E-04& 2.1E+07& n\\
&&&&&\\
&  4.81& 0.000& 0.078&1.70E+08&
 0.772& 1.414&   4.070&4.18E+08&
HeCO   & 4.3E-06& 7.4E-05& 7.4E+05& n\\
&&&&&\\
&  9.85& 0.000& 0.102&1.70E+08&
 0.790& 1.411&   7.972&3.48E+08&
HeCO   & 5.4E-06& 1.0E-04& 5.9E+05& n\\
&&&&&\\
& 19.35& 0.000& 0.112&1.70E+08&
 0.803& 1.410&  15.121&2.41E+09&
HeCO?  & 2.8E-06& 1.0E-04& 5.0E+05& n\\
&&&&&&&&1.28&&  34.9& 3.165& 1.4007&  9.13\\
\tablebreak
\tablevspace{0pt}
\multicolumn{14}{l}{$M_2^i =  4.5\Ms$:}\\
\tablevspace{6pt}
&  0.57& 0.682& 0.000&1.26E+06&
 3.145& 1.403&   0.155&2.44E+06&
Del Dyn& 3.5E-07& 1.0E-04& 1.3E+06& n\\
&&&&&&&&1.50&1235.6&  89.9& 4.001& 1.4025&  0.33\\
&  1.20& 0.342& 0.000&8.45E+07&
 2.686& 1.402&   0.241&1.77E+06&
Del Dyn& 9.0E-07& 1.0E-04& 7.9E+05& n\\
&&&&&&&&1.50& 731.3&  42.6& 4.427& 1.4017&  1.11\\
&  2.38& 0.103& 0.000&1.14E+08&
 0.471& 1.507&   3.432&3.73E+08&
HeCO?  & 9.5E-08& 1.0E-04& 1.9E+07& y\\
&&&&&&&&1.47& 319.1& 100.1& 4.453& 1.4015&  2.26\\
&  4.98& 0.000& 0.084&1.22E+08&
 0.889& 1.409&   1.856&2.99E+09&
HeCO?  & 1.6E-05& 1.0E-04& 6.9E+04& y\\
&&&&&&&&1.41&&   9.4& 4.494& 1.4001&  4.95\\
& 14.08& 0.000& 0.113&1.22E+08&
 2.675& 1.400&   2.816&1.02E+06&
Dyn    & 5.3E-05& 1.0E-04& 1.3E+04& n\\
&&&&&&&&1.50&&   8.6& 4.159& 1.4001&  9.79\\
& 38.60& 0.000& 0.134&1.22E+08&
 4.005& 1.400&  22.625&1.01E+06&
Dyn    & 9.1E-06& 1.0E-04& 1.9E+03& n\\
&&&&&&&&1.50&&   1.9& 4.490& 1.4000& 38.25\\
\tablevspace{0pt}
\multicolumn{14}{l}{$M_2^i =  5.0\Ms$:}\\
\tablevspace{6pt}
&  0.59& 0.677& 0.000&2.20E+06&
 3.609& 1.401&   0.131&1.56E+06&
Del Dyn& 8.3E-07& 1.0E-04& 5.3E+05& n\\
&&&&&&&&1.50& 500.9&  24.5& 4.519& 1.4012&  0.34\\
&  1.22& 0.338& 0.000&6.82E+07&
 3.544& 1.401&   0.254&1.34E+06&
Del Dyn& 1.9E-06& 1.0E-04& 3.1E+05& n\\
&&&&&&&&1.50& 291.7&  17.6& 4.360& 1.4011&  0.58\\
&  2.42& 0.101& 0.000&9.10E+07&
 3.377& 1.401&   0.436&1.21E+06&
Del Dyn& 3.1E-06& 1.0E-04& 1.9E+05& n\\
&&&&&&&&1.50& 164.8&  20.9& 4.604& 1.4008&  1.51\\
&  4.61& 0.000& 0.081&9.72E+07&
 3.553& 1.400&   0.965&1.02E+06&
Dyn    & 2.3E-05& 1.0E-04& 8.7E+03& n\\
&&&&&&&&1.50&&   8.7& 4.763& 1.4001&  3.46\\
&  9.50& 0.000& 0.114&9.73E+07&
 4.035& 1.400&   3.174&1.01E+06&
Dyn    & 2.7E-05& 1.0E-04& 5.3E+03& n\\
&&&&&&&&1.43&&   5.3& 4.805& 1.4001&  7.51\\
& 19.30& 0.000& 0.124&9.73E+07&
 4.006& 1.400&   6.246&1.01E+06&
Dyn    & 3.2E-05& 1.0E-04& 4.0E+03& n\\
&&&&&&&&1.50&&   4.0& 4.864& 1.4001& 16.32\\
& 37.90& 0.000& 0.133&9.73E+07&
 4.160& 1.400&  14.408&1.02E+06&
Dyn    & 2.2E-05& 1.0E-04& 9.5E+03& n\\
&&&&&&&&1.50&&   9.5& 4.781& 1.4002& 29.08\\
\tablebreak
\tablevspace{0pt}
\multicolumn{14}{l}{$M_2^i =  6.0\Ms$:}\\
\tablevspace{6pt}
&  0.62& 0.666& 0.000&3.51E+06&
 5.004& 1.401&   0.170&1.29E+06&
Del Dyn& 7.5E-07& 1.0E-04& 2.7E+05& n\\
&&&&&&&&1.50& 262.5&   8.3& 5.773& 1.4006&  0.46\\
&  1.25& 0.338& 0.000&4.42E+07&
 5.035& 1.401&   0.355&1.20E+06&
Del Dyn& 8.2E-07& 1.0E-04& 1.8E+05& n\\
&&&&&&&&1.50& 174.3&   6.7& 5.841& 1.4006&  1.01\\
&  2.46& 0.104& 0.000&5.90E+07&
 5.226& 1.401&   0.886&1.15E+06&
Del Dyn& 1.5E-06& 1.0E-04& 1.3E+05& n\\
&&&&&&&&1.50& 124.6&   8.5& 5.853& 1.4005&  2.02\\
&  4.97& 0.000& 0.094&6.31E+07&
 5.362& 1.400&   2.131&1.01E+06&
Dyn    & 2.2E-05& 1.0E-04& 3.7E+03& n\\
&&&&&&&&1.50&&   3.7& 5.911& 1.4001&  4.40\\
&  9.94& 0.000& 0.123&6.32E+07&
 5.448& 1.400&   4.749&1.01E+06&
Dyn    & 2.8E-05& 1.0E-04& 2.4E+03& n\\
&&&&&&&&1.50&&   2.4& 5.928& 1.4000&  9.01\\
& 21.21& 0.000& 0.140&6.32E+07&
 5.542& 1.400&  11.471&1.01E+06&
Dyn    & 3.3E-05& 1.0E-04& 1.8E+03& n\\
&&&&&&&&1.50&&   1.8& 5.934& 1.4000& 19.39\\
& 41.27& 0.000& 0.145&6.32E+07&
 5.570& 1.400&  23.152&1.01E+06&
Dyn    & 3.5E-05& 1.0E-04& 1.4E+03& n\\
&&&&&&&&1.50&&   1.4& 5.943& 1.4000& 38.07\\
\tablevspace{0pt}
\multicolumn{14}{l}{$M_2^i =  7.0\Ms$:}\\
\tablevspace{6pt}
&  0.64& 0.668& 0.000&2.30E+06&
 6.322& 1.401&   0.241&1.22E+06&
Del Dyn& 4.2E-07& 1.0E-04& 2.0E+05& n\\
&&&&&&&&1.50& 193.6&   6.8& 6.945& 1.4004&  0.60\\
&  1.25& 0.349& 0.000&3.10E+07&
 6.395& 1.401&   0.522&1.15E+06&
Del Dyn& 2.9E-07& 1.0E-04& 1.4E+05& n\\
&&&&&&&&1.50& 138.8&   2.8& 6.933& 1.4004&  1.15\\
&  2.56& 0.102& 0.000&4.23E+07&
 6.421& 1.400&   1.106&1.10E+06&
Del Dyn& 5.0E-07& 1.0E-04& 9.2E+04& n\\
&&&&&&&&1.50&  90.3&   2.2& 6.933& 1.4004&  2.32\\
&  5.27& 0.000& 0.108&4.52E+07&
 6.528& 1.400&   2.646&1.01E+06&
Dyn    & 2.5E-05& 1.0E-04& 1.7E+03& n\\
&&&&&&&&1.50&&   1.7& 6.952& 1.4000&  4.90\\
& 10.88& 0.000& 0.131&4.52E+07&
 6.585& 1.400&   5.937&1.01E+06&
Dyn    & 2.8E-05& 1.0E-04& 1.3E+03& n\\
&&&&&&&&1.50&&   1.3& 6.958& 1.4000& 10.22\\
& 21.46& 0.000& 0.147&4.52E+07&
 6.654& 1.400&  12.960&1.00E+06&
Dyn    & 3.2E-05& 1.0E-04& 1.0E+03& n\\
&&&&&&&&1.50&&   1.0& 6.963& 1.4000& 20.30\\
& 42.23& 0.000& 0.152&4.52E+07&
 6.714& 1.400&  27.771&1.00E+06&
Dyn    & 3.3E-05& 1.0E-04& 8.2E+02& n\\
&&&&&&&&1.50&&   0.8& 6.968& 1.4000& 40.23\\

\enddata
\tablecomments{{\it First row:} $P^i\,$: initial orbital 
period (d); $X_c^i$: initial
central hydrogen mass fraction; $M_c^i/M$: initial fractional mass of the H-exhausted
core; $t^i\,$: age at beginning of mass transfer (yr); $M_2^f$, $M_1^f$:
final masses of the secondary and primary ($M_\odot$); $P^f\,$: final
orbital period (d); $\Delta t\,$: total time since beginning of mass transfer
(yr); Type: type of evolution: `short': compact system; `He': He white dwarf
secondary; `HeCO': HeCO white dwarf secondary; `del dyn': delayed dynamical
instability; `dyn': dynamical mass transfer; `?' indicates that the system
may be dynamically unstable;  $<\dot{M}>$: average mass-transfer
rate ($M_{\odot}\,$yr$^{-1}$); $\dot{M}_{\rm max}$: maximum mass-transfer
rate ($M_{\odot}\,$yr$^{-1}$); $\Delta t_{\dot{M}}\,$: duration of
mass-transfer phase (yr); H flash?: occurrence of H shell flashes \\
\rule{0pt}{12pt}
{\it Second row:} period gap (hr); $M_2^{\rm min}$: secondary mass
at period minimum ($M_\odot$); $P_{\rm min}\,$: minimum period (min);
$X_s^{\rm min}$: surface H abundance at period minimum; 
$f_{\rm over}$: maximum overflow factor; $t_{\rm atm}$: duration
of atmospheric Roche-lobe overflow phase (kyr); $t_{\rm rad}$: duration
of radiative Roche-lobe overflow phase (kyr);  $M_2^{\rm dyn}$, $M_1^{\rm 
dyn}$, $P^{\rm dyn}$: secondary and primary mass ($M_\odot$), orbital
period (d) at $\dot{M}=10^{-4}\, M_\odot\,$yr$^{-1}$.\\
$^{a}$: $M_{\odot}\,$yr$^{-1}$}
\end{deluxetable}



\begin{references}

\reference{Alexander} Alexander, D. R., \& Ferguson, J. W. 1994, ApJ, 437, 879

\reference{Anderson} Anderson, S. B. 1992, Ph.D. Thesis, California
Institute of
Technology

\reference{Bailyn} Bailyn, C. D., \& Grindlay, J. E. 1987, ApJ, 316, L25

\reference{Baraffe} Baraffe, I., Chabrier, G., Allard, F. \& Hauschildt, \&
P. H. 1998, A\&A, 337, 403

\reference{Bhatta} Bhattacharya, D., \& van den Heuvel, E. P. J. 1991,
Phys.\ Rep., 203, 1

\reference{Blundell} Blundell, K. M., Mioduszewski, A. J., Muxlow, T. W. B.,
Podsiadlowski, Ph., Rupen, M. P. 2001, ApJ, submitted

\reference{Braun} Braun, H. 1997, Ph.D. Thesis (LMU, Munich)

\reference{Camilo} Camilo, F., Lorimer, D. R., Freire, P., Lyne, A. G.,
\& Manchester, R. N. 2000, ApJ, 535, 975

\reference{Casares} Casares, J., Charles, P., \& Kuulkers, E. 1998, ApJ,
493, L39

\reference{Chakrabarty} Chakrabarty, D. 1998, ApJ, 492, 342

\reference{Chou} Chou, Y., Grindlay, J. E., \& Bloser, P. F. 2001, ApJ, 549,
1135

\reference{Davies} Davies, M. B., \& Hansen, B. M. S. 1998, MNRAS, 301, 15

\reference{Deutsch} Deutsch, E. W., Margon, B., \& Anderson, S. F. 2000, ApJ,
530, L21

\reference{Distefano}  Di\,\,Stefano, R., Nelson, L. A., Lee, W., Wood, T. H.,
Rappaport, S. 1997, in Thermonuclear Supernovae, NATO ASI Series,
ed.\ P. Ruiz-Lapuente, R. Canal, \&  J. Isern (Dordrecht: Kluwer), 147

\reference{Distefano2} Di\,\,Stefano, R., \& Rappaport, S. 1992, ApJ,
396, 587

\reference{Eggleton2} Eggleton, P. P. 1983, ApJ, 268, 368

\reference{Eggleton} Eggleton, P. P., Faulkner, J., \& Flannery, B. P. 1973,
A\&A, 23, 325

\reference{Ergma1} Ergma, E. 1996, A\&A, 315, L17

\reference{Ergma2} Ergma, E., \& Sarna, M. J. 1996 MNRAS, 280, 1000

\reference{Fabian} Fabian, A. C., Pringle, J. E., Rees, M. J. 1975, MNRAS,
172, 15

\reference{Faulkner} Faulkner, 1971, ApJ, 170, 99

\reference{Fedorova} Fedorova, A. V., \& Ergma, E. V. 1989, Ap\&SS, 151, 125

\reference{Finger} Finger, M. H., Koh, D. T., Nelson, R. W.,
 Prince, T. A., Vaughan, B. A., \& Wilson, R. B. 1996, Nat, 381, 291

\reference{Fruchter2} Fruchter, A. S., \& Goss, W. M. 1990, ApJ, 365, 63

\reference{Fruchter} Fruchter, A. S., Stinebring, D. R., \& Taylor, J. H.
 1988, Nat, 333, 237


\reference{Hameury} Hameury, J. M., King, A. R., Lasota, J. P., \&
 Raison, F. 1993, A\&A, 277, 81

\reference{Han2} Han, Z., Tout, C. A., \& Eggleton, P. P. 2000, MNRAS,
319, 215

\reference{Harpaz} Harpaz, A., \& Rappaport, S. 1991, ApJ, 383, 739

\reference{Hjellming} Hjellming, M. S., \& Webbink, R. F. 1987, ApJ, 318, 794

\reference{Homer} Homer, L.,  Charles, P. A., Naylor, T., van Paradijs, J.,
Auriere, M., \& Koch-Miramond, L. 1996, MNRAS, 282, L37

\reference{Howell} Howell, S. B., Nelson, L. A., \& Rappaport, S. 2001,
ApJ, 550, 898

\reference{Iben} Iben, I., Jr., \& Tutukov, A. V. 1985, ApJS, 58, 661

\reference{Iben2} Iben, I., Jr., \& Tutukov, A. V. 1986, ApJ, 311, 742

\reference{Ilovaisky} Ilovaisky, S. A. et al. 1993, A\&A, 270, 139

\reference{In't} in't Zand, J. J. M. et al.\ 2000, A\&A, 355, 145

\reference{Johnston} Johnston, S., \& Bailes, M. 1991, MNRAS, 252, 277

\reference{Joss} Joss, P. C., \& Rappaport, S, 1984, ARA\&A, 22, 537

\reference{King2} King, A. R., \& Begelman, M. C. 1999, ApJ, 519, L169

\reference{King4} King, A. R., Kolb, U., \& Sienkiewicz, E. 1997, ApJ,
488, 89

\reference{King3} King, A. R., Schenker, K., Kolb, U., \& Davies, M. B.
2001, MNRAS, 321, 327

\reference{King1} King, A. R., \& Ritter, H. 1999, MNRAS, 309, 253


\reference{Kippenhahn2} Kippenhahn, R., Thomas, H.-C., \& Weigert,
A. 1968, Z. Astrophys., 68, 256

\reference{Kippenhahn3} Kippenhahn, R., \& Weigert, A. 1990, Stellar
Structure and
Evolution (Berlin: Springer)

\reference{Kippenhahn} Kippenhahn, R., Weigert, A., \& Hofmeister, E. 1967, in
Methods in Computational Physics, Vol. 7, ed.\ B. Alder, S. Fernbach, \&
M. Rothenberg (New York: Academic), 129


\reference{Kolb2} Kolb, U., \& Baraffe, I. 1999, MNRAS, 309, 1034

\reference{Kolb} Kolb, U., Davies, M. B., King, A., \& Ritter, H. 2000, MNRAS,
317, 438

\reference{Kulkarni} Kulkarni, S. R., \& Narayan, R. 1988, ApJ, 335, 755

\reference{Kulkarni2} Kulkarni, S, R., Narayan, R., \& Romani, R. W.
1990, ApJ, 356, 174.

\reference{Landau} Landau, L. D., Lifshitz, E. M. 1959, The Classical
Theory of Fields (Pergamon Press: Oxford)

\reference{Langer} Langer, N., Deutschmann, A., Wellstein, S., \& H\"oflich, P.
2000, A\&A, 361, 1046

\reference{Lewin} Lewin, W. H. G.,
van Paradijs, J., \& van den Heuvel, E. P. J. (eds.) 1995, X-Ray Binaries
(Cambridge, Cambridge Univ.\ Press)

\reference{McMillan} McMillan, S. L. W., McDermott, P. N., \& Taam, R. E.
 1987, ApJ, 318, 261

\reference{Middleditch} Middleditch, J., Mason, K. O.,
 Nelson, J. E., White, N. E. 1981, ApJ, 244, 1001

\reference{Nelson} Nelson, L. A., Rappaport, S. A., \& Joss, P. C.
1986, ApJ, 304, 231

\reference{Orosz} Orosz, J. A., \& Kuulkers, E. 1999, MNRAS, 305, 132

\reference{Paczynski2} Paczy\'nski, B. 1970, in IAU Colloq.\ 6, Mass Loss
and Evolution in Close Binaries, ed.\ K. Gyldenkerne \& R. M. West
(Copenhagen: Copenhagen University Publications), 139


\reference{Paczynski3} Paczy\'nski, B. 1976, in Structure and Evolution
in Close Binary Systems, ed.\ P. P. Eggleton, S. Mitton, \& J. Whelan
(Dordrecht: Reidel), 75

\reference{Paczynski4} Paczy\'nski, B., \& Sienkiewicz,
R.  1972, Acta Astron., 22, 73

\reference{Paczynski} Paczy\'nski, B., \& Sienkiewicz, R. 1981,
ApJ,  248, L27

\reference{Pennington} Pennington, R. P. 1986, Ph.D.\ Thesis
(Cambridge University), unpublished

\reference{Pfahl} Pfahl, E. D., Podsiadlowski, Ph., \& Rappaport, S. 2001,
in preparation

\reference{Podsi3} Podsiadlowski, Ph.\ 1991, Nat, 350, 136

\reference{Podsi2} Podsiadlowski, Ph.\ 1996, MNRAS, 279, 1104

\reference{Podsi1} Podsiadlowski, Ph., \& Rappaport, S. 2000, ApJ, 529, 946

\reference{Pols} Pols, O. R., Tout, C. A., Schr\"oder, K.-P.,
Eggleton, P. P., \& Manners, J. 1997, MNRAS, 289, 869

\reference{Press2} Press, W. H., \& Teukolsky, S. A. 1977, ApJ, 213, 183

\reference{Press} Press, W. H., Teukolsky, S. A., Vetterling, W. T.,
Flannery, B. P. 1992,  Numerical Recipes in Fortran: the Art
of Scientific Computing (Cambridge: Cambridge University Pres), Chapter 9.3

\reference{Pylyser1} Pylyser, E. H. P., \& Savonije, G., J. 1988, A\&A, 191, 57

\reference{Pylyser2} Pylyser, E. H. P., \& Savonije, G., J. 1989, A\&A, 208, 52

\reference{Rappaport4} Rappaport, S., Di\,\,Stefano, R., \& Smith, M. 1994,
ApJ, 426, 692

\reference{Rappaport6} Rappaport, S., \& Joss, P. C. 1997, ApJ, 486, 435

\reference{Rappaport3} Rappaport, S. Joss, P. C., \& Webbink, R. F. 1982,
ApJ, 254, 616

\reference{Rappaport5} Rappaport, S., Pfahl, E., Rasio, F.,
\& Podsiadlowski, Ph.\ 2001, in Evolution of Binary and Multiple Star Systems,
ASP Conf.\ Proc., Vol.~229, ed.\ Ph.\ Podsiadlowski, S. Rappaport,
A. R. King, F. D'Antona, \& L. Burderi (San Francisco, ASP), 409

\reference{Rappaport1} Rappaport, S., Podsiadlowski, Ph., Joss, P. C.,
Di\,\,Stefano, R., \& Han, Z. 1995, MNRAS, 273, 731

\reference{Rappaport2} Rappaport, S., Verbunt, F., \& Joss, P. C. 1983,
ApJ, 275, 713


\reference{Rasio} Rasio, F. A., Pfahl, E. D., \& Rappaport, S. 2000, ApJ,
532, L47

\reference{Rasio2} Rasio, F. A., \& Shapiro, S. L., 1991, ApJ, 377, 559

\reference{Ray} Ray, A.,  Kembhavi, A. K., \& Antia, H. M. 1987, A\&A,
184, 164

\reference{Ritter} Ritter, H. 1988, A\&A, 202, 93

\reference{Ritter3} Ritter, H. 1996, in Evolutionary Processes in Binary
Stars, ed.\ R. A. M. J. Wijers, M. B. Davies, \& C. A. Tout (Dordrecht:
Kluwer), 223

\reference{Ritter2} Ritter, H., \& Kolb, U. 1998, A\&AS, 129, 83

\reference{Rogers} Rogers, F. J., Iglesias, C. A. 1992, ApJS, 79, 507

\reference{Ruderman1} Ruderman, M., Shaham, J., \& Tavani, M.
1989, ApJ, 336, 507

\reference{Ruderman2} Ruderman, M., Shaham, J., Tavani, M., \&
Eichler, D. 1989, ApJ, 343, 292

\reference{Sansom} Sansom, A. E. Dotani, T., Asai, K., \& Lehto, H. J.
1993, MNRAS, 262, 429

\reference{Sarna} Sarna, M. J., Ergma, E., \& Ger\v skevit\v s-Antipova, J.
2000, MNRAS, 316, 84

\reference{Savonije} Savonije, G. J., de Kool, M., \& van den Heuvel,
E. P. J. 1986, A\&A, 155, 51

\reference{Sawada} Sawada, E., Hachisu, I., \& Matsuda, T. 1984, MNRAS,
206, 673

\reference{Schroder} Schr\"oder, K.-P., Pols, O. R., Eggleton, P. P. 1997,
MNRAS, 285, 696

\reference{Soberman} Soberman, G. E., Phinney, E. S., \& van den
Heuvel, E. P. J., 1997, A\&A, 327, 620

\reference{Spruit} Spruit, H. C., \& Ritter, H. 1983, A\&A, 124, 267

\reference{Stella} Stella, L., White, N., \& Priedhorsky, W. 1987, ApJ,
315, L49

\reference{Swank} Christian, D. J., \& Swank, J. H. 1997, ApJS, 109, 177

\reference{Taam} Taam, R. E., \& van den Heuvel, E. P. J. 1986, ApJ, 302, 235.

\reference{Tan} Tan, J., et al. 1991, ApJ, 374, 291

\reference{Tauris1} Tauris, T. M., \& Savonije, G. J. 1999, A\&A, 350, 928

\reference{Tauris2} Tauris, T. M., van den Heuvel, E. P. J., \& Savonije,
G. J. 2000, ApJ, 530, L93

\reference{Taylor} Taylor, J. H., Manchester, R. N., \& Lyne, A. G. 1993,
ApJS, 88, 529.

\reference{Thomas} Thomas, H.-C. 1967, Z.\ Astrophys., 67, 420

\reference{Thorne} Thorne, K. S., \& \.Zytkow, A. N. 1977, ApJ, 212, 832

\reference{Tutukov} Tutukov, A. V., Fedorova, A. V., Ergma E., \& Yungelson,
L. R. 1985, SvA, 11, 123

\reference{Tutukov2} Tutukov, A. V., Fedorova A. V., Ergma E.,
\& Yungelson L. 1987, Sov. Astron. Lett., 13, 328

\reference{Heuvel} van den Heuvel, E. P. J., \& van Paradijs, J. 1988,
Nat, 334, 227.

\reference{Klis} van der Klis, M., et al.\ 1993, MNRAS, 260, 686


\reference{Paradijs2} van Paradijs, J. 1995, in X-Ray Binaries,
ed.\ W. H. G. Lewin, J. van Paradijs, \& E. P. J. van den Heuvel
(Cambridge: Cambridge University Press), 536

\reference{Paradijs} van Paradijs, J. 1996, ApJ, 464, L139

\reference{Verbunt2} Verbunt, F. 1987, ApJ, 312, 23

\reference{Verbunt} Verbunt, F., \& Zwaan, C. 1981, A\&A, 100, L7

\reference{Witte} Witte, M. G., \& Savonije, G. J. 2001, A\&A, 366, 840

\end{references}
\end{document}